\documentclass[9pt,twocolumn]{extarticle}

\usepackage{graphicx}
\usepackage{amssymb}
\usepackage{amsmath}
\usepackage[english]{babel}
\usepackage{color}
\usepackage{cite}

\begin{document}

\title{\bf  Binding equilibrium and kinetics of membrane-anchored receptors and ligands in cell adhesion: insights from computational model systems and theory}

\author{Thomas R.\ Weikl$^{1}$,  Jinglei Hu$^{1,2}$, Guang-Kui Xu$^{1,3}$, and Reinhard Lipowsky$^{1}$\\
\small $^1$ Max Planck Institute of Colloids and Interfaces, Department of Theory and Bio-Systems, 14424 Potsdam, Germany\\
\small $^2$ Kuang Yaming Honors School, Nanjing University, 210023 Nanjing, China\\
\small $^3$ International Center for Applied Mechanics, State Key Laboratory for Strength and Vibration of Mechanical Structures, \\[-2pt]
\small  Xi'an Jiaotong University, Xi'an 710049, China}

\date{}

\maketitle

\begin{abstract}
The adhesion of cell membranes is mediated by the binding of membrane-anchored receptor and ligand proteins. In this article, we review recent results from simulations and theory that lead to novel insights on how the binding equilibrium and kinetics of these proteins is affected by the membranes and by the membrane anchoring and molecular properties of the proteins. Simulations and theory both indicate that the binding equilibrium constant $K_{\rm 2D}$ and the on- and off-rate constants of anchored receptors and ligands in their `two-dimensional' (2D) membrane environment strongly depend on the membrane roughness from thermally excited shape fluctuations on nanoscales. Recent theory corroborated by simulations provides a general relation between $K_{\rm 2D}$ and the binding constant $K_{\rm 3D}$ of soluble variants of the receptors and ligands that lack the membrane anchors and are free to diffuse in three dimensions (3D). 
\end{abstract}
%

\section{Introduction}

Cell adhesion processes and the adhesion of vesicles to the membranes of cells or organelles depend sensitively on the binding constant and binding kinetics of the membrane-anchored receptor and ligand molecules that mediate adhesion. Since the binding equilibrium constant $K_{\rm 2D}$ and the on- and off-rate constants of these receptor and ligand molecules are difficult to measure in their natural two-dimensional (2D) membrane environment, a central question is how they are related to the binding equilibrium constant $K_{\rm 3D}$ and the on- and off-rate constants of soluble variants of the receptors and ligands that lack the membrane anchors and are free to diffuse in three dimensions (3D)
\cite{Dustin01,Orsello01,Krobath09,Wu11,Leckband12,Zarnitsyna12,Hu13,Wu13,Xu15,Hu15}. The binding constant $K_{\rm 3D}$ and on- and off-rate constants of these soluble receptors and ligands can be quantified with standard experimental methods \cite{Schuck97,Rich00,McDonnell01}. 

The binding equilibrium constant $K_{\rm 2D}$ of membrane-anchored receptor and ligand molecules has units of area, while the binding constant $K_{\rm 3D}$ of soluble variants of these molecules has units of volume. Bell and co-workers \cite{Bell84} therefore suggested the relation $K_{\rm 2D} = K_{\rm 3D}/l_c$ between the binding constants with a characteristic confinement length $l_c$ that balances the different units of these constants. However, experimental data for $K_{\rm 2D}$ and $K_{\rm 3D}$ of several receptor and ligand pairs lead to values of the confinement length $l_c$ that can differ by orders of magnitude, depending on whether $K_{\rm 2D}$ is determined with fluorescence methods  or with mechanical methods \cite{Dustin01}. Fluorescence methods \cite{Dustin96,Dustin97,Zhu07,Tolentino08,Huppa10,Axmann12,ODonoghue13} probe the binding equilibrium of receptors and ligands in equilibrated adhesion zones of cells and lead to values of $l_c$ of the order of nanometers. In contrast, mechanical methods \cite{Kaplanski93,Alon95,Piper98,Chesla98,Merkel99,Williams01,Chen08,Chien08,Huang10,Liu14} probe the binding kinetics of anchored receptors and ligands during initial contacts and typically lead to values of $l_c$ between tens of micrometers and millimeters in cell adhesion experiments \cite{Dustin01}.

In this article, we review recent results from computational model systems and theory that provide general and novel insights 
into the relation between the binding equilibrium and kinetics of membrane-anchored receptor and ligand molecules in 2D and the binding of soluble variants of these molecules in 3D. A central aspect of these computational and theoretical results is that the relation between the binding equilibrium constants $K_{\rm 2D}$ and $K_{\rm 3D}$ involves four characteristic lengths, rather than a single confinement length \cite{Xu15}. Two of these four lengths are characteristic lengths of the receptor-ligand complex that reflect variations in the binding site, and how strongly the local membrane separation at the location of the complex is constrained by the complex. The remaining two lengths are the average separation and relative roughness of the apposing membranes and, thus, characteristic lengths of the membranes. The relative membrane roughness is the local standard deviation of the membranes from their average separation due to thermally excited shape fluctuations on nanoscales. 

The binding equilibrium constant $K_{\rm 2D}$ strongly depends both on the average membrane separation and the relative membrane roughness, which helps to understand why mechanical methods that probe the binding kinetics of membrane-anchored proteins during initial membrane contacts can lead to values for $K_{\rm 2D}$ that are orders of magnitude smaller than the values obtained from fluorescence measurements in equilibrated adhesion zones \cite{Xu15}.  In equilibrated adhesion zones that are dominated by a single species of receptors and ligands, the average membrane separation is close to the preferred average separation for receptor-ligand binding at which $K_{\rm 2D}$ is maximal, and the relative membrane roughness is reduced by receptor-ligand bonds \cite{Krobath09,Hu13}. During initial membrane contacts, in contrast, both the average membrane separation and relative membrane roughness are larger, which can lead to significantly smaller values of $K_{\rm 2D}$.

\section{Characteristic lengths of membranes and membrane-anchored receptors and ligands} 

A membrane-anchored receptor can only bind to an apposing membrane-anchored ligand if the local membrane separation $l$ at the site of the receptor and ligand is within an appropriate range. This local separation $l$ of the membranes varies -- along the membranes, and in time -- because of thermally excited membrane shape fluctuations. Experiments that probe the binding equilibrium constant $K_{\rm 2D}$ or the on- and off-rate constants $k_{\rm on}$ and $k_{\rm off}$ imply averages in space and time over membrane adhesion regions and measurement durations. Our recent simulations and theories indicate that these averages  can be expressed as \cite{Xu15,Hu15}
\begin{align}
K_{\rm 2D} &= \int K_{\rm 2D}(l) P(l) {\rm d}l 
\label{K2Dav} \\
k_{\rm on} &= \int k_{\rm on}(l) P(l) {\rm d}l
\label{konav}
\end{align}
where $K_{\rm 2D}(l)$ and $k_{\rm on}(l)$ are the binding equilibrium constant and on-rate constant as functions of the local membrane separation $l$, and $P(l)$ is the distribution of local membrane separations that reflects the spatial and temporal variations of $l$. The single-peaked functions $K_{\rm 2D}(l)$ and $k_{\rm on}(l)$ are maximal the at preferred local separation of the receptors and ligands for binding, and have characteristic widths that depend on the anchoring, length, and flexibility of the receptors and ligands \cite{Xu15,Hu15}. The off-rate constant follows from Eqs.\ (\ref{K2Dav}) and (\ref{konav}) as $k_{\rm off}=k_{\rm on}/K_{\rm 2D}$. Our simulations also show that the distribution $P(l)$ of the local  separation is well approximated by the Gaussian distribution 
\begin{equation}
P(l) \simeq \exp\left[-(l-\bar l)^2/2\xi_\perp^2\right]/(\sqrt{2\pi} \xi_\perp)
\label{Pl}
\end{equation}
in situations in which the adhesion of two apposing membranes, or membrane segments, is mediated by a single type of receptors and ligands \cite{Hu15,Xu15}. Here, $\bar{l}=\langle l \rangle$ is the average separation of the membranes or membrane segments, and $\xi_\perp = \sqrt{\langle(l - \bar{l})^2\rangle}$ is the relative roughness of the membranes. The relative roughness is the standard deviation of the local membrane separation $l$, i.e.~the width of the distribution $P(l)$.
The distribution $P(l)$ describes both the spatial and temporal variations of the local membrane separation $l$ of two apposing membranes, or membrane segments. Related temporal averages for the on-rate constant $k_{\rm on}$ and off-rate constant $k_{\rm off}$ at fixed membrane locations have been employed by Bihr et al.\ \cite{Bihr12}.

The Eqs.\ (\ref{K2Dav}) and (\ref{Pl}) illustrate three characteristic lengths of the binding constant $K_{\rm 2D}$. These lengths are the width $\xi_{\rm RL}$ of the single-peaked function $K_{\rm 2D}(l)$, which reflects how strongly the local separation $l$ is constrained by a receptor-ligand (RL) complex, and the average separation $\bar{l}$ and relative roughness $\xi_\perp$ of the membranes. A fourth characteristic length that affects the relation of the binding constants $K_{\rm 2D}$ and $K_{\rm 3D}$ in our theory is the ratio $V_b/A_b$ of the translational space phase volume $V_b$ of a bound soluble receptor in 3D and the translational phase space area $A_b$ of a bound membrane-anchored receptor in 2D, relative to their ligands (see Section IV). Similarly, three characteristic lengths of the on-rate constant $k_{\rm on}$ are the width $\xi_{\rm TS}$ of the single-peaked function $k_{\rm on}(l)$, which reflects variations of the local separation $l$ in the transition-state (TS) complex for binding, the average membrane separation $\bar{l}$, and the relative membrane roughness $\xi_\perp$, according to Eqs.\ (\ref{konav}) and (\ref{Pl}).

\begin{figure*}[htp]
\begin{center}
\resizebox{1.6\columnwidth}{!}{\includegraphics{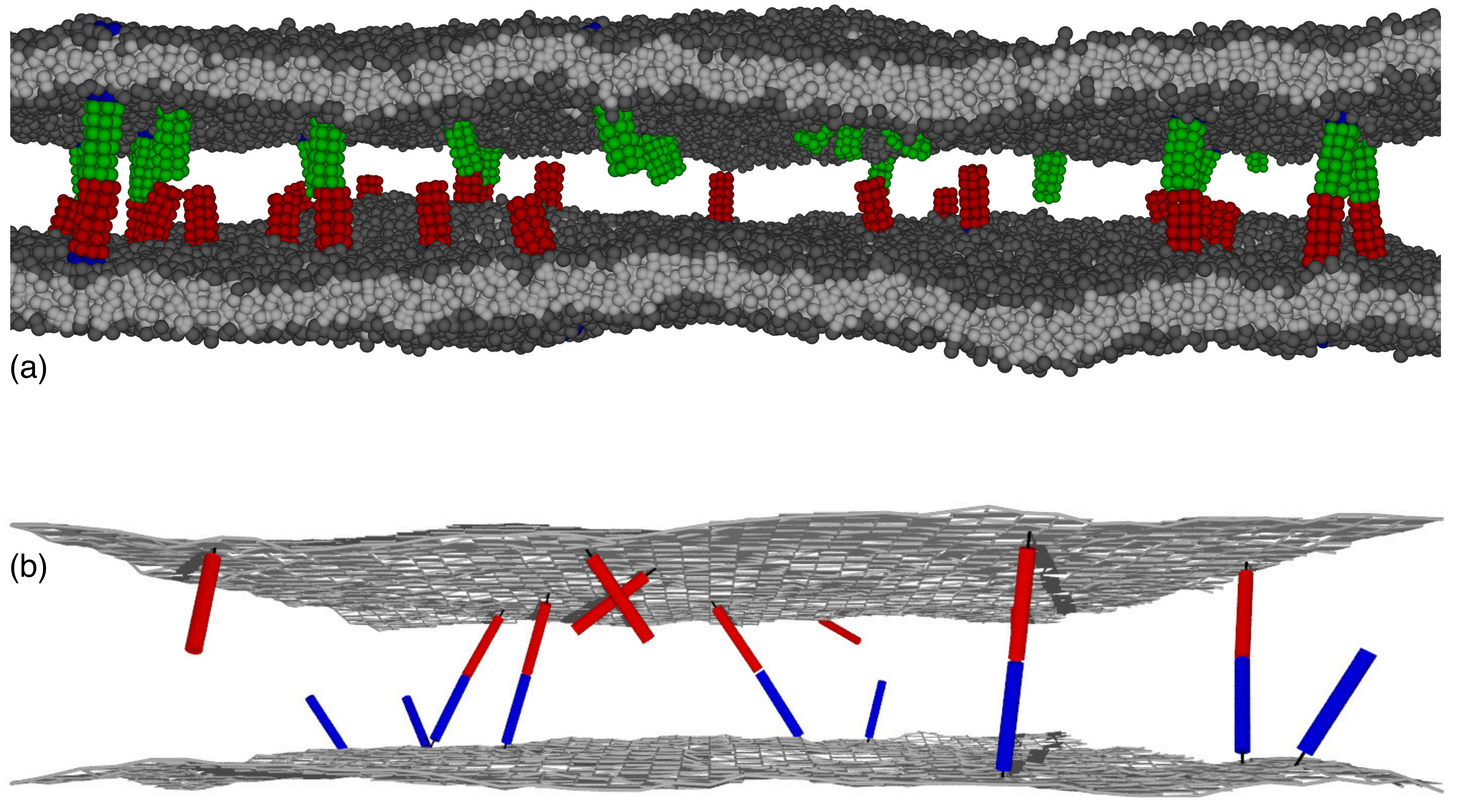}}
\end{center}
\caption{ 
(a) Snapshot from a molecular dynamics (MD) simulation of our coarse-grained molecular model of biomembrane adhesion. In this snapshot, the two apposing membranes both have an area of $120\times 120$ ${\rm nm}^2$ and contain 25 transmembrane receptors and ligands.
(b) Snapshot from a Monte Carlo (MC) simulation of our elastic-membrane model of biomembrane adhesion. The snapshot shows membrane segments of area $200 \times 200$ nm$^2$ from simulations with overall membrane area $800 \times 800$ nm$^2$ and 200 receptors and ligands of anchoring strength $k_a = 4 \, k_B T$ and length 20 nm. 
}
\label{figure-snapshots}
\end{figure*}

In equilibrated membrane adhesion zones that are dominated by a single type of receptors and ligands, the average membrane separation is close to the preferred average separation of these receptors and ligands for binding. Our simulations indicate that the relative membrane roughness $\xi_\perp$ then is determined by the concentration $[{\rm RL}]$ of the receptor-ligand bonds, which constrain the membrane shape fluctuations \cite{Xu15,Krobath07}:
\begin{equation}
\xi_\perp \simeq 0.2 \sqrt{(k_B T / \kappa_{\rm eff})} \Big /
\sqrt{[{\rm RL}]}
\label{relative_roughness_scaling}
\end{equation}
Here, $\kappa_{\rm eff} = \kappa_1\kappa_2/(\kappa_1 + \kappa_2)$ is the effective bending rigidity of the two apposing membranes with bending rigidities $\kappa_1$ and $\kappa_2$, and $k_B T$ is the thermal energy, the driving force of membrane shape fluctuations. For a concentration $[{\rm RL}]\simeq 100/\mu{\rm m}^{2}$ of receptor-ligand bonds and for typical values of the bending rigidities $\kappa_1$ and $\kappa_2$ of lipid membranes \cite{Nagle13,Dimova14} and cell membranes \cite{Pontes13,Betz09} between 20 $k_B T$ and 80 $k_B T$, we obtain estimates for the relative membrane roughness $\xi_\perp$ between 3 nm and 6 nm from Eq.\ (\ref{relative_roughness_scaling}). For a four times larger bond concentration $[{\rm RL}]\simeq 400/\mu{\rm m}^{2}$, these roughness estimates are decreased by a factor of 2, according to Eq.\ (\ref{relative_roughness_scaling}). For a four times smaller bond concentration $[{\rm RL}]\simeq 25/\mu{\rm m}^{2}$, the roughness estimates are increased by a factor of 2, compared to the bond concentration $[{\rm RL}]\simeq 100/\mu{\rm m}^{2}$. The scaling relation (\ref{relative_roughness_scaling}) results from the fact that the membrane shape fluctuations on the relevant lateral length scales 
up to $1/\sqrt{[{\rm RL}]}$, i.e.\ on length scales of the order of 10 or 100 nanometers, are dominated by the bending energy of the membranes. In contrast, the overall shape of cells on length scales of micrometers is dominated by the membrane tension  and the cell cytoskeleton. The bending energy dominates over the membrane tension $\sigma$ on length scales smaller than the crossover length $\sqrt{\kappa/\sigma}$, which adopts values of 100 or a few 100 nanometers for typical values of the bending rigidity $\kappa$ and tension $\sigma$ of cell membranes \cite{Pontes13}. 

If the relative membrane roughness $\xi_\perp$ is much smaller than the widths $\xi_{\rm RL}$ and $\xi_{\rm TS}$ of the functions $K_{\rm 2D}(l)$ and $k_{\rm on}(l)$, the binding of membrane-anchored receptors and ligands is only weakly affected by $\xi_\perp$. Such situations may occur in focal contacts or adherens junctions, which consist of clusters of integrin and cadherin complexes, respectively \cite{Zamir01,Wu11,Leckband14,Biswas15,Yap15}. In cell adhesion zones of immune cells and in the equilibrated adhesion zones probed with fluorescence methods  \cite{Dustin96,Dustin97,Zhu07,Tolentino08,Huppa10,Axmann12,ODonoghue13}, in contrast, the relative membrane roughness is likely of the same order or larger than $\xi_{\rm RL}$ and $\xi_{\rm TS}$. The computational model systems and theory described in the next sections indicate that the binding equilibrium and kinetics of the membrane-anchored receptors and ligands is then strongly affected both by the relative membrane roughness $\xi_\perp$ and the average membrane separation $\bar{l}$. If the relative membrane roughness $\xi_\perp$ is significantly larger than $\xi_{\rm RL}$ and $\xi_{\rm TS}$, the binding equilibrium constant $K_{\rm 2D}$ and on-rate constant $k_{\rm on}$ are both inversely proportional to $\xi_\perp$ at the preferred average separation for binding \cite{Xu15,Hu15}. Together with Eq. (\ref{relative_roughness_scaling}), these inverse proportionalities lead to a quadratic dependence of the bond concentration $[{\rm RL}]$ and the overall reaction rate on the concentrations $[{\rm R}]$ and $[{\rm L}]$ of unbound membrane-anchored receptors R and ligands L, which reflects the binding cooperativity caused by the membrane roughness on nanoscales \cite{Hu15,Hu13,Krobath09}.

\section{Results from computational model systems of biomembrane adhesion}

We have recently developed two computational model systems to investigate the binding of anchored receptors and ligands in their 2D membrane environment and the binding of soluble variants of the receptors and ligands that are fully mobile in 3D \cite{Hu13,Xu15,Hu15}. First, we have developed a coarse-grained molecular model of biomembrane adhesion \cite{Hu13,Hu15} (see Fig.\ \ref{figure-snapshots}(a)). In this model, lipid molecules consist of three hydrophobic head beads and two hydrophobic tails of four beads each, and the receptors and ligands are represented as cylindrical rods of beads, which are either anchored rather rigidly to a cylindrical transmembrane domain, or more flexibly to  lipid molecules. We have investigated the binding equilibrium and kinetics of both these transmembrane and lipid-anchored receptors and ligands with molecular dynamics (MD) simulations, as well as the binding equilibrium and kinetics of soluble variants of the receptors and ligands that lack the membrane anchors. Related coarse-grained molecular models of biomembranes have been previously used to investigate the self-assembly \cite{Goetz98,Shelley01,Marrink04,Shih06}, fusion \cite{Marrink03,Shillcock05,Grafmuller07,Grafmuller09,Smirnova10,Risselada11}, 
and lipid domains \cite{Illya06,Risselada08,Meyer10,Apajalahti10,Bennett13} of membranes as well as the diffusion \cite{Gambin06,Guigas06}, aggregation \cite{Reynwar07}, and curvature generation \cite{Arkhipov08,Simunovic13} of membrane proteins with MD simulations.

Second, we have developed an elastic-membrane model of biomembrane adhesion in which the membranes are represented as discretized elastic surfaces, and the receptors and ligands as anchored rigid or semi-flexible rods that diffuse continuously along the membranes and rotate around their anchoring points \cite{Xu15}. Using Monte Carlo (MC) simulations, we have determined both the binding constant $K_{\rm 2D}$ of these anchored receptors and ligands as well as the binding constant  $K_{\rm 3D}$ of soluble variants of the receptors and ligands. In previous elastic-membrane models of biomembrane adhesion, determining both $K_{\rm 2D}$ and $K_{\rm 3D}$ and the molecular characteristics affecting these binding constants has not been possible because the receptors and ligands are not explicitly represented as anchored molecules. Instead, the binding of receptors and ligands has been described implicitly by interactions that depend on the membrane separation \cite{Lipowsky96,Weikl01,Weikl02a,Weikl04,Asfaw06,Tsourkas07,Reister08,Paszek09,Bihr12}. In other previous elastic-membrane models, receptors and ligands are described by concentration fields rather than individual molecules \cite{Komura00,Bruinsma00,Qi01,Chen03,Raychaudhuri03,Coombs04,Shenoy05,Wu06,Zhang08a,Atilgan09}, or receptor-ligand bonds are treated as constraints on the local membrane separation \cite{Zuckerman95,Krobath07,Speck10,Weil10,Dharan15}.

An important aspect for the binding of membrane-anchored receptors and ligands is the flexibility of the membrane anchoring. In our computational model systems, the anchoring flexibility of unbound membrane-anchored receptors and ligands can be described by the harmonic anchoring energy
\begin{equation}
V_{\rm anchor}=\frac {1}{2} k_a \theta_a^2
\label{Vanchor}
\end{equation}
with anchoring strength $k_a$ and anchoring angle $\theta_a$, which is the angle between the direction of the receptors and ligands and the local membrane normal. An anchoring angle of zero thus corresponds to a perpendicular orientation of the receptors and ligands relative to the membrane. For our coarse-grained molecular model of biomembrane adhesion, the effective anchoring strength $k_a$ can be determined by fitting the anchoring-angle distributions of unbound receptors and ligands observed in the MD simulations, which leads to the values $k_a \simeq 2.5$ $k_B T$ for our lipid-anchored receptors and ligands and $k_a \simeq 23$ $k_B T$ for our transmembrane receptors and ligands \cite{Hu15}. In our elastic-membrane model of biomembrane adhesion, the anchoring energy (\ref{Vanchor}) of receptors and ligands is part of the overall configurational energy of the model, and the anchoring strength $k_a$ thus can be `set' as a parameter.  We have performed MC simulations with the three values $k_a = 4$, $8$ and $16$ $k_B T$.  

\begin{figure}[htp]
\resizebox{\columnwidth}{!}{\includegraphics{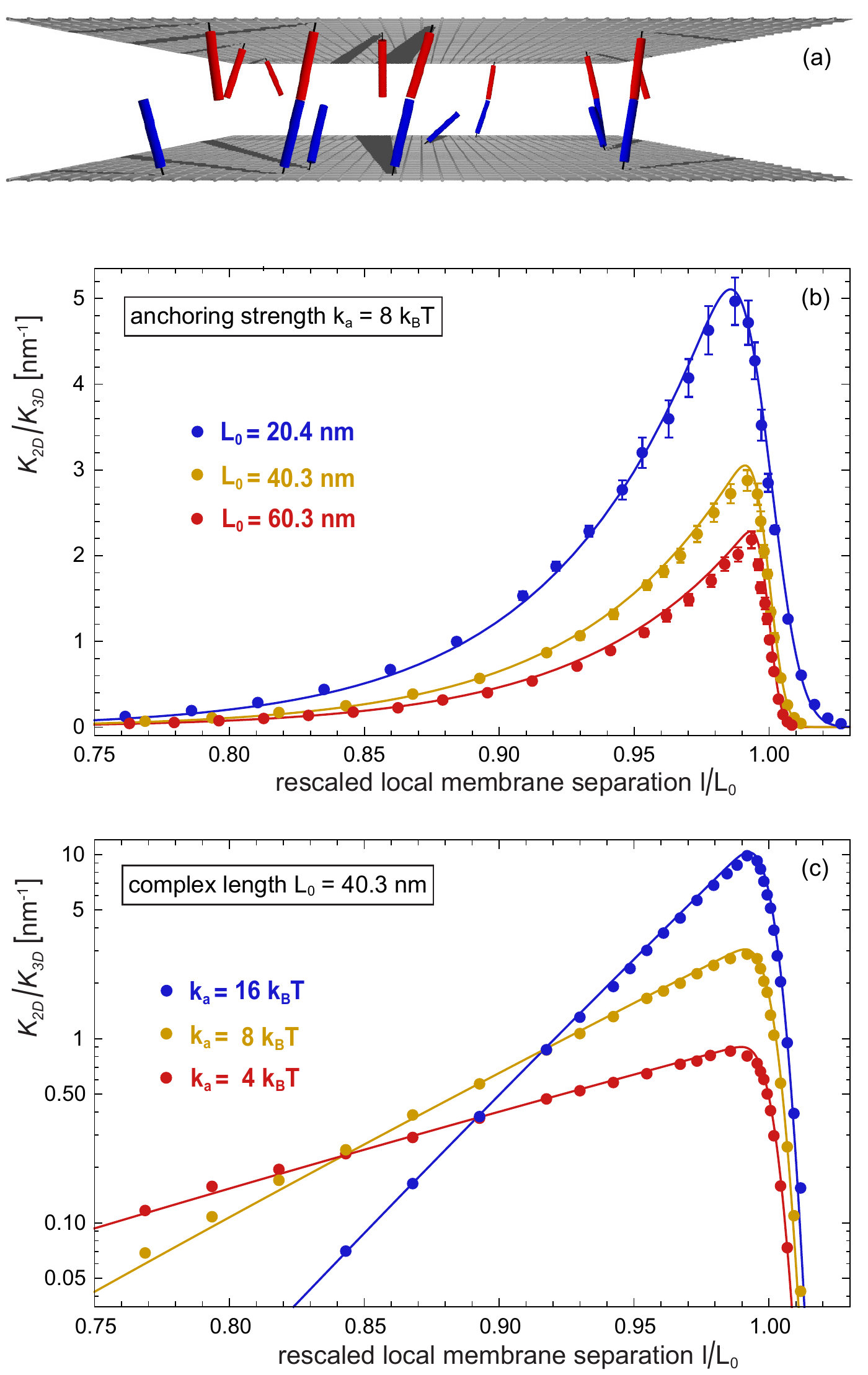}}
\caption{
(a) Snapshot from a MC simulation with parallel and planar membranes. (b) and (c) Ratio $K_{\rm 2D}/K_{\rm 3D}$ of the binding constants of membrane-anchored and soluble receptors and ligands versus local membrane separation $l$ for different anchoring strengths $k_a$ and complex lengths $L_0$ of the receptors and ligands of our elastic-membrane model of biomembrane adhesion. The data points represent MC data, and the lines theoretical results based on Eqs.\ (\ref{HRL}) and (\ref{K2Dl-B}). The binding constant $K_{\rm 3D}$ of soluble variants of the receptors and ligand is determined by the binding potential of the receptors and ligands and does not depend on the complex length $L_0$.
}
\label{figure-MC-planar}
\end{figure}
\begin{figure}[htp]
\resizebox{\columnwidth}{!}{\includegraphics{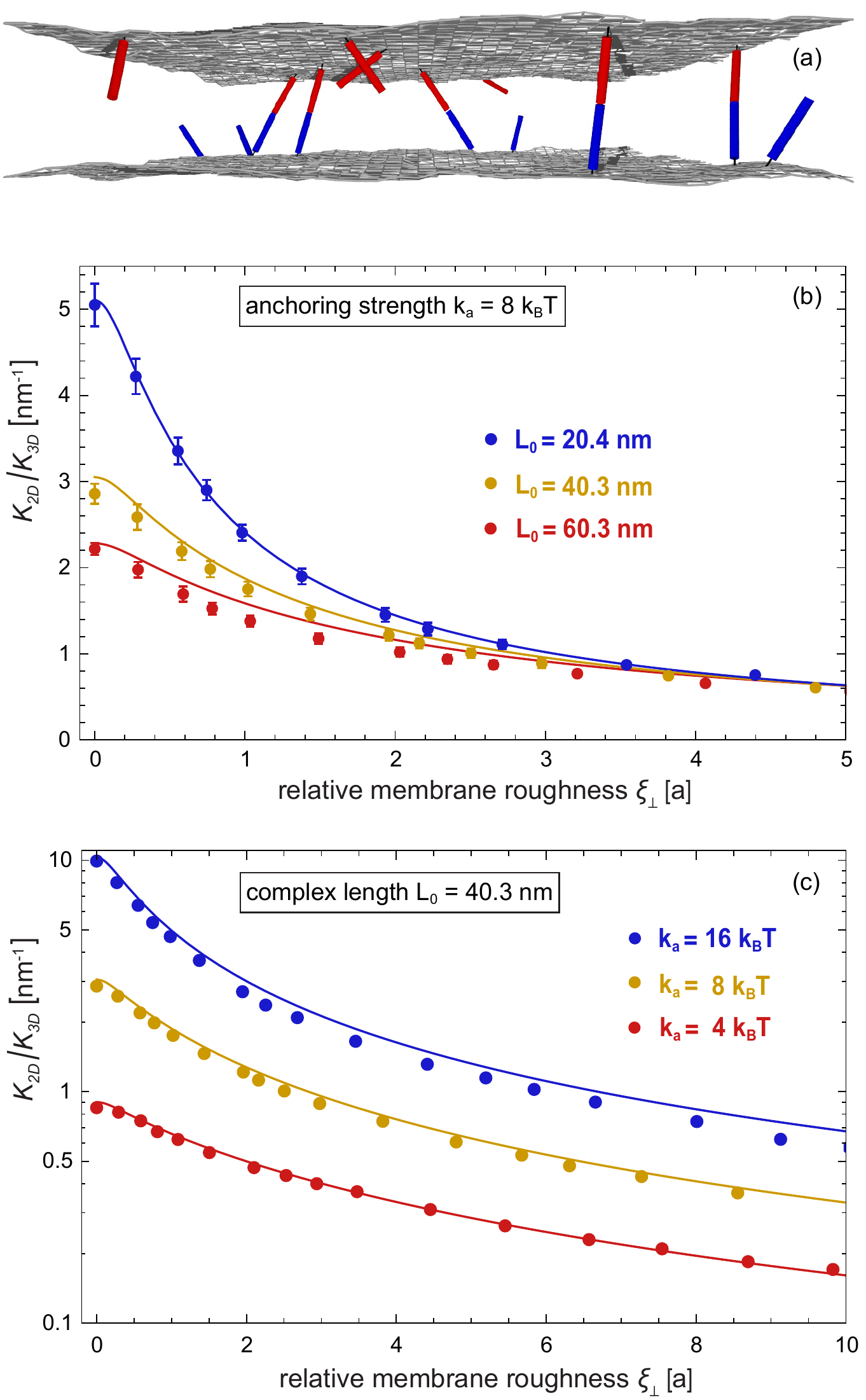}}
\caption{
(a) Snapshot from a MC simulation with fluctuating membranes.
(b) and (c) Ratio $K_{\rm 2D}/K_{\rm 3D}$ of the binding constants of membrane-anchored and soluble receptors and ligands versus 
relative membrane roughness $\xi_\perp$ of two equilibrated fluctuating membranes with preferred average separation for different anchoring strengths $k_a$ and complex lengths $L_0$ of the receptors and ligands.  The data points represent MC data, and the lines represent theoretical results based on Eqs.\ (\ref{K2Dav}),  (\ref{HRL}), and (\ref{K2Dl-B}).
}
\label{figure-MC-fluc}
\end{figure}

The Figs.\  \ref{figure-MC-planar} and \ref{figure-MC-fluc} illustrate MC results for the binding constant of membrane-anchored receptors and ligands 
 from two different simulation scenarios \cite{Xu15}. In the first scenario, the two apposing membranes are parallel and planar (see Fig.\ \ref{figure-MC-planar}(a)). The local separation $l$ of the membranes is then identical at all membrane sites and,  thus, identical to the average separation $\bar{l}$ of the membranes. By varying the membrane separation $l$ in this scenario, we obtain the binding constant $K_{\rm 2D}$ as a function of the local membrane separation $l$ from MC simulations in which the receptors and ligands diffuse along the planar membranes and rotate at their anchor points. In the second scenario, the two apposing membranes are flexible, and the local membrane separation $l$ varies because of thermally excited shape fluctuations of the membranes (see Fig.\ \ref{figure-MC-fluc}(a)). These variations can be quantified by the relative roughness $\xi_\perp$ of the membranes, which is the standard deviation of the local separation. In this scenario, the membranes are `free to choose'  an optimal average separation $\bar{l}_0$ at which the overall free energy is minimal, and we obtain $K_{\rm 2D}$ as a function of the membrane roughness $\xi_\perp$ at the average membrane separation $\bar{l}=\bar{l}_0$ from MC simulations that differ in the numbers of receptors and ligands, and in the membrane tension.  In both MC simulations scenarios, the binding constant of the membrane-anchored receptors and ligands is obtained as $K_{\rm 2D} = [{\rm RL}]_{\rm 2D}/[{\rm R}]_{\rm 2D}[{\rm L}]_{\rm 2D}$ from the average area concentrations $[{\rm RL}]_{\rm 2D}$,  $[{\rm R}]_{\rm 2D}$, and $[{\rm L}]_{\rm 2D}$ of the bound receptor-ligand complexes, unbound receptors, and unbound ligands observed in the simulations. The binding constant of soluble variants of the receptors and ligands can be obtained as $K_{\rm 3D} = [{\rm RL}]_{\rm 3D}/[{\rm R}]_{\rm 3D}[{\rm L}]_{\rm 3D}$ from the volume concentrations of the receptors and ligands observed in MC simulations. The binding constant $K_{\rm 3D}$ is determined by the binding potential of our model, and does not depend on the length of the complexes \cite{Xu15}.

\begin{figure*}[htp]
\resizebox{2\columnwidth}{!}{\includegraphics{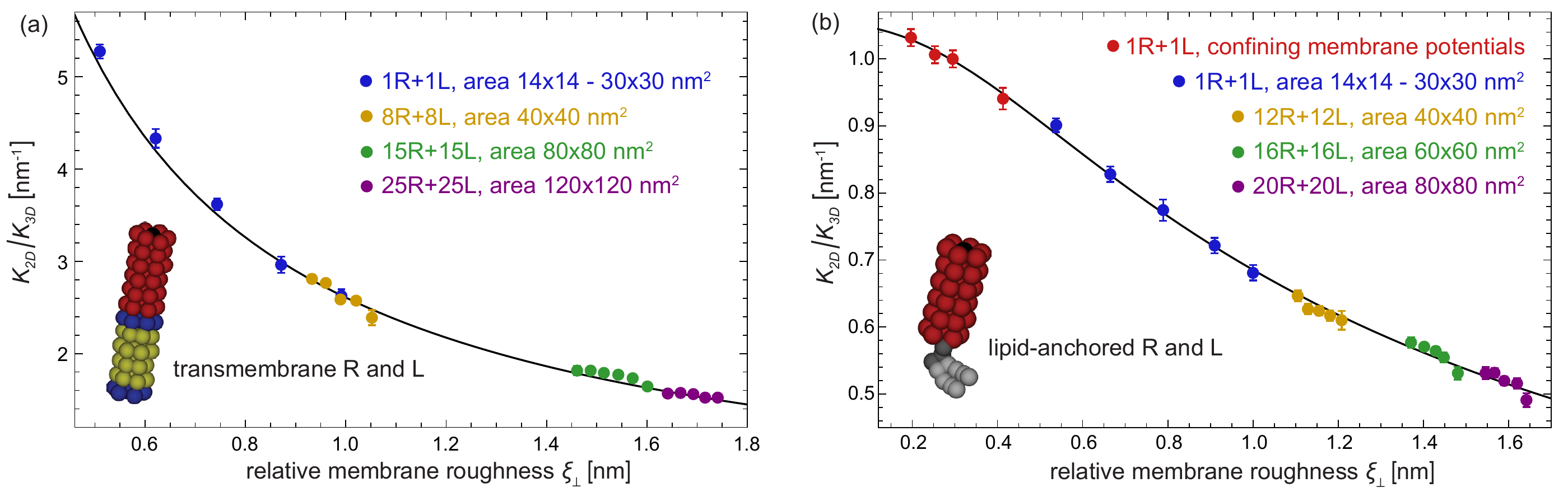}}
\caption{\small Ratio $K_{\rm 2D}/K_{\rm 3D}$ of the binding constants of membrane-anchored and soluble receptors and ligands versus 
relative membrane roughness $\xi_\perp$ at the preferred average separation for (a) transmembrane and (b) lipid-anchored receptors and ligands of our coarse-grained molecular model of biomembrane adhesion. The MD data points result from a variety of membrane systems. In these systems, the area of the two apposing membranes ranges from $14 \times 14$ nm$^2$ to $120 \times 120$ nm$^2$, and the number of receptors (R) and ligands (L) varies between 1 and 25 (see figure legends). For membrane systems with several receptors and ligands, we obtain multiple data points for states that differ in the number of bound receptor-ligand complexes \cite{Hu13,Hu15}. The red data points in (b) result from simulations with confining membrane potentials that restrict membrane shape fluctuations. In experiments, such a situation occurs for membranes bound to apposing surfaces as, e.g., in the surface force apparatus \cite{Israelachvili92,Bayas07}. 
The full line in (a) represents a fit to Eq.\ (\ref{K2Dlim}) for the average membrane separation $\bar{l} = \bar{l}_0$ with fit parameter $\tilde{c}_{\rm 2D}= 2.6\pm 0.2$. We fit to Eq.\ (\ref{K2Dlim}) because the characteristic length $\xi_{\rm RL}$ of our transmembrane receptors and ligands is about 0.38 nm and thus smaller than the values of the relative membrane roughness of all membrane systems in (a). The full line in (b) results from a fit based on Eqs.\ (\ref{K2Dav}), (\ref{HRL}), and (\ref{K2Dl-B}) with fit parameters $c_{\rm 2D}=420\pm 40$ nm$^2$,   $L_0 = 10.35\pm 0.05$ nm, and $k_{\rm RL} = 6.0 \pm 1.0$ $k_BT/{\rm nm}^2$ for the anchoring strength $k_a \simeq 2.5$ $k_BT$ of lipid-anchored receptors and ligands obtained from the anchoring-angle distributions of the unbound receptors and ligands.
}
\label{figure-MD-optsep}
\end{figure*}

As a function of the local separation $l$, the binding constant $K_{\rm 2D}(l)$ is maximal at a local membrane separation $l_0$ that is slightly smaller than the length $L_0$ of the receptor-ligand complexes, and is asymmetric with respect to $l_0$  (see Fig.\ \ref{figure-MC-planar}(b) and (c)). This asymmetry reflects that the receptor-ligand complexes can tilt at local separations $l$ smaller than $l_0$, but need to stretch at local separations larger than $l_0$. The maximum of the function $K_{\rm 2D}(l)$ decreases with increasing length $L_0$ of the rigid receptor-ligand complexes (see Fig.\ \ref{figure-MC-planar}(b)), and strongly increases with increasing anchoring strength $k_a$ of the receptors and ligands (see Fig.\ \ref{figure-MC-planar}(c)). The width of the function $K_{\rm 2D}(l)$ increases with decreasing anchoring strength $k_a$. These features of the function $K_{\rm 2D}(l)$ can be understood from our general theory presented in the next section, which agrees with the MC data without any fit parameters (see full lines in Fig.\ \ref{figure-MC-planar}). 

The MC data in Fig.\  \ref{figure-MC-fluc} and the corresponding MD data of Fig.\ \ref{figure-MD-optsep} illustrate that the binding constant $K_{\rm 2D}$ of receptors and ligands anchored to fluctuating membranes decreases with increasing relative membrane roughness $\xi_\perp$ at the optimal average membrane separation $\bar{l}_0$ for binding. In Fig.\ \ref{figure-MC-fluc}, the ratio $K_{\rm 2D}/K_{\rm 3D}$ of the binding constant, the inverse `confinement length', varies between 0.2 and 10 nm$^{-1}$, depending on the relative roughness $\xi_\perp$ of the membranes and on the anchoring strength and length of the receptors and ligands. 

In Fig.\  \ref{figure-MD-optsep}, the values of $K_{\rm 2D}/K_{\rm 3D}$ range from 0.5 to 5 nm$^{-1}$, depending on the relative membrane roughness  $\xi_\perp$ and on whether the receptors and ligands have a transmembrane anchor or a lipid anchor. The MD data points in Fig.\ \ref{figure-MD-optsep} result from a variety of membrane systems that differ in membrane area, in the number of receptors and ligands, or in the membrane potential \cite{Hu15}. The roughness depends on the area $L_x \times L_y$ of the membranes in the MD simulations because the periodic boundaries of the simulation box suppress membrane shape fluctuations with wavelength larger than $L_x/2 \pi$ where $L_x = L_y$ is the linear membrane size. In membrane systems with several anchored receptors and ligands, the roughness is affected by the number of receptor-ligand bonds because the bonds constrain the membrane shape fluctuations. For the small numbers of receptors and ligands in our MD simulations, the binding constants can be determined from the times spent in bound and unbound states \cite{Hu13,Hu15}.

The binding kinetics of the transmembrane and lipid-anchored receptors and ligands of our coarse-grained molecular model of biomembrane adhesion can be determined from the frequencies of binding and unbinding events observed in MD simulations \cite{Hu13}. The binding potential is identical for both types of receptors and ligands and has no barrier to ensure an efficient sampling of binding and unbinding events of receptors and ligands in our simulations. The kinetics of these events is then strongly enhanced compared with protein binding events in experiments \cite{Huppa10,Huang10,Robert11,Axmann12,ODonoghue13}. However, this rate enhancement does not affect our main results, which concern the dependence of the rate constants and equilibrium constant on the membrane separation and roughness. At the preferred average separation $\bar{l}_0$ for binding, the 2D on-rates of the anchored receptors and ligands {\em decrease} with the relative membrane roughness, while the 2D off-rates {\em increase} with the relative roughness \cite{Hu13,Hu15}. For our transmembrane receptors and ligands, the 2D off-rate $k_{\rm off}$ increases from about $90/ {\rm ms}$ to about $140/ {\rm ms}$ with an increase of the relative membrane roughness from 0.5 nm to 1.8 nm for the membrane systems of Fig.\ \ref{figure-MD-optsep}(a). For our lipid-anchored receptors and ligands, the 2D off-rate $k_{\rm off}$ increases from about $245/ {\rm ms}$ to about $290/ {\rm ms}$ with an increase of the relative membrane roughness from 0.2 nm to 1.7 nm for the membrane systems of Fig.\ \ref{figure-MD-optsep}(b). The 3D off-rate of soluble variants of these receptors and ligands with the same binding potential is $k_{\rm off} \simeq 400 / {\rm ms}$. This 3D off-rate is slightly larger than the off-rates of the lipid-anchored receptors and ligands, and about 3 to 5 times larger than the off-rates of the transmembrane receptors and ligands at the preferred average separation for binding. These results appear to indicate that the 2D off-rates of the receptors and ligands in our coarse-grained molecular model are smaller than the 3D off-rate due to constraints on the rotational motion from membrane anchoring, which are more pronounced for our transmembrane receptors and ligands. 2D off-rates that are slightly smaller than 3D off-rates have also been observed for the binding of T-cell receptors to MHC-peptides in experiments in which the T-cell cytoskeleton is disrupted \cite{Huppa10}. In experiments with intact T-cell cytoskeleton, the 2D off-rates are affected by ATP-driven cytoskeletal forces exerted on TCR-MHC-peptide complexes \cite{Huppa10,Huang10,ODonoghue13,Liu14,Wang12}.

\section{General theory for the binding equilibrium and kinetics of membrane-anchored receptors and ligands} 

We have derived a general theory for the binding equilibrium and kinetics of membrane-anchored receptors and ligands that agrees with the results from our computational model systems. In this theory, the binding constants $K_{\rm 2D}$ and $K_{\rm 3D}$ of membrane-anchored and soluble receptors and ligands can be calculated from the translational and rotational free-energy change upon binding. As a function of the local membrane separation $l$, the binding constant $K_{\rm 2D}$ has the general form \cite{Xu15}
\begin{equation}
K_{\rm 2D}(l) \simeq  \sqrt{8\pi} K_{\rm 3D} \frac{A_b}{V_b} \frac{\Omega_{\rm RL}(l)}{\Omega_{\rm R}\Omega_{\rm L}}
\label{K2Dl-A}
\end{equation}
in this theory. Here,  $\Omega_{\rm R}$, $\Omega_{\rm L}$, and $\Omega_{\rm RL}(l)$ are the rotational phase space volumes of the unbound receptors R, unbound ligands L, and bound receptor-ligand complex RL relative to the membranes, and $A_b$ and $V_b$ are the translational phase space area and translational phase space volume of the bound ligand relative to the receptor in 2D and 3D. The ratio $V_b/A_b$ in Eq.\ (\ref{K2Dl-A}) represents a characteristic length for the binding interface of the receptor-ligand complex and can be estimated as the standard deviation of the binding-site distance in the direction of the complex  \cite{Xu15}.  The rotational phase space volumes of the unbound receptors and ligands can be calculated as 
$\Omega_{\rm R} = \Omega_{\rm L} =  2\pi \int_0^{\pi/2} \exp[-\frac{1}{2} k_a \theta_a^2/k_B T] \sin \theta_a \, {\rm d}\theta_a$. The remaining, theoretically `challenging' term in Eq.\ (\ref{K2Dl-A}) is the rotational phase space volume $\Omega_{\rm RL}(l)$ of the bound complex, which determines the shape of the function $K_{\rm 2D}(l)$. 

We have found that the rotational phase space volume $\Omega_{\rm RL}(l)$ of the bound receptor-ligand complex can be calculated from an effective configurational energy $H_{\rm RL}$ of the bound receptor-ligand complex. In our computational model systems, the binding angles and binding angle variations of the rigid, rod-like receptor and ligand molecules are small compared to their anchoring-angle variations. A receptor and ligand then have an approximately collinear orientation in the complex, and approximately equal anchoring angles $\theta_a$. The effective configurational energy is then
\begin{align}
H_{\rm RL}(l,\theta_a) \simeq k_a \theta_a^2 + \frac{1}{2} k_{\rm RL}(l /\! \cos \theta_a - L_0)^2 
\label{HRL}
\end{align}
The first term of this effective energy is the sum of the anchoring energies ({\ref{Vanchor}}) for the receptor and ligand in the complex, and the second term is a harmonic approximation for variations in the length $L_{\rm RL}$ of the receptor-ligand complex, i.e.~in the distance between the two anchoring points of the complex. For parallel membranes with separation $l$ and approximately identical anchoring angles $\theta_a$ of the RL complex in these membranes, the length of the complex, i.e.\ the distance between the two anchoring points in the membranes, is $L_{\rm RL} \simeq l /\! \cos \theta_a$. With the effective configurational energy (\ref{HRL}), the rotational phase space volume of the bound complex can be calculated as 
$\Omega_{\rm RL} \simeq  2\pi \int_0^{\pi/2} \exp[-H_{\rm ef}/k_B T] \sin \theta_a \, {\rm d}\theta_a$, which leads to 
\begin{equation}
K_{\rm 2D}(l) = 2\pi  c_{\rm 2D} \int_0^{\pi/2} e^{-H_{\rm RL}(l,\theta_a)/k_B T}  \sin\theta_a {\rm d}\theta_a
\label{K2Dl-B}
\end{equation}
with $c_{\rm 2D} = \sqrt{8\pi} K_{\rm 3D} A_b / (V_b \Omega_{\rm R}\Omega_{\rm L})$.

The theoretical result for $K_{\rm 2D}(l)$ of Eq.\ (\ref{K2Dl-B}) agrees with MC data for our elastic-membrane model of biomembrane adhesion without any fit parameters (see lines in Fig.\ \ref{figure-MC-planar}). For our elastic-membrane model, the effective spring constant $k_{\rm RL}$ and preferred length $L_0$ of the receptor-ligand complex in the effective configurational energy (\ref{HRL}) can be calculated from the standard deviations of the binding angle and binding-site distance and from the lengths of the receptors and ligands. By combining the Eqs.\ (\ref{K2Dav}), (\ref{Pl}), and (\ref{K2Dl-B}), we obtain general results for the binding constant $K_{\rm 2D}$ of receptors and ligands anchored to fluctuating membranes that agree with MC data without fit parameters (see lines in Fig.\ \ref{figure-MC-fluc}). Our general theory for the binding constant $K_{\rm 2D}$ thus captures the essential features of the `dimensionality reduction' from 3D to 2D due to membrane anchoring.

In analogy to Eq.\  (\ref{HRL}) for the bound receptor-ligand complex, we have postulated the effective configurational energy
\begin{align}
H_{\rm TS}(l,\theta_a) \simeq k_a \theta_a^2 + \frac{1}{2} k_{\rm TS}(l /\! \cos \theta_a - L_{TS})^2 
\label{HTS}
\end{align}
for the transition-state complex of the binding reaction of membrane-anchored receptors and ligands, with the same anchoring strength $k_a$ as in Eq.\  (\ref{HRL}). This effective configurational energy reflects that a receptor and ligand molecule can only bind at appropriate relative orientations and separations. The effective spring constant $k_{\rm TS}$ for the length variations of the transition-state complex is smaller than the corresponding spring constant $k_{\rm RL}$ of the RL complex, because the variations in the binding-site distance and binding angle, which affect the effective spring constants, are larger in the transition state \cite{Hu15}. The preferred effective length $L_{\rm TS}$ of the transition-state complex, in contrast, is in general close to the preferred length $L_0$ of the bound RL complex. In analogy to Eq.\ (\ref{K2Dl-B}), the on-rate constant is 
\begin{equation}
k_{\rm on}(l) \simeq  2 \pi c_{\rm on} \int_0^{\pi/2}  e^{-H_{\rm TS}(l,\theta_a)/k_B T}  \sin\theta_a {\rm d}\theta_a
\label{konl}
\end{equation}
for a given separation $l$ of the planar and parallel membranes. The integration over the angle $\theta_a$ in Eq.\ (\ref{konl}) can be interpreted as an integration over the transition-state ensemble of the binding reaction. The on-rate constant $k_{\rm on}$ of receptors and ligands anchored to fluctuating membranes can then be obtained from an average over the local membrane separation $l$ (see Eq. (\ref{konav})). This average over local separations for the on-rate constant $k_{\rm on}$ relies on characteristic timescales for membrane fluctuations that are significantly smaller than the timescales for the diffusion of the anchored molecules on the relevant length scales \cite{Hu15,Bihr12}. In contrast, the average in Eq.\ (\ref{K2Dav}) for the binding constant $K_{\rm 2D}$ is independent of these timescales because $K_{\rm 2D}$ is an equilibrium quantity that does not depend on dynamic aspects.

\begin{figure}[htp]
\resizebox{\columnwidth}{!}{\includegraphics{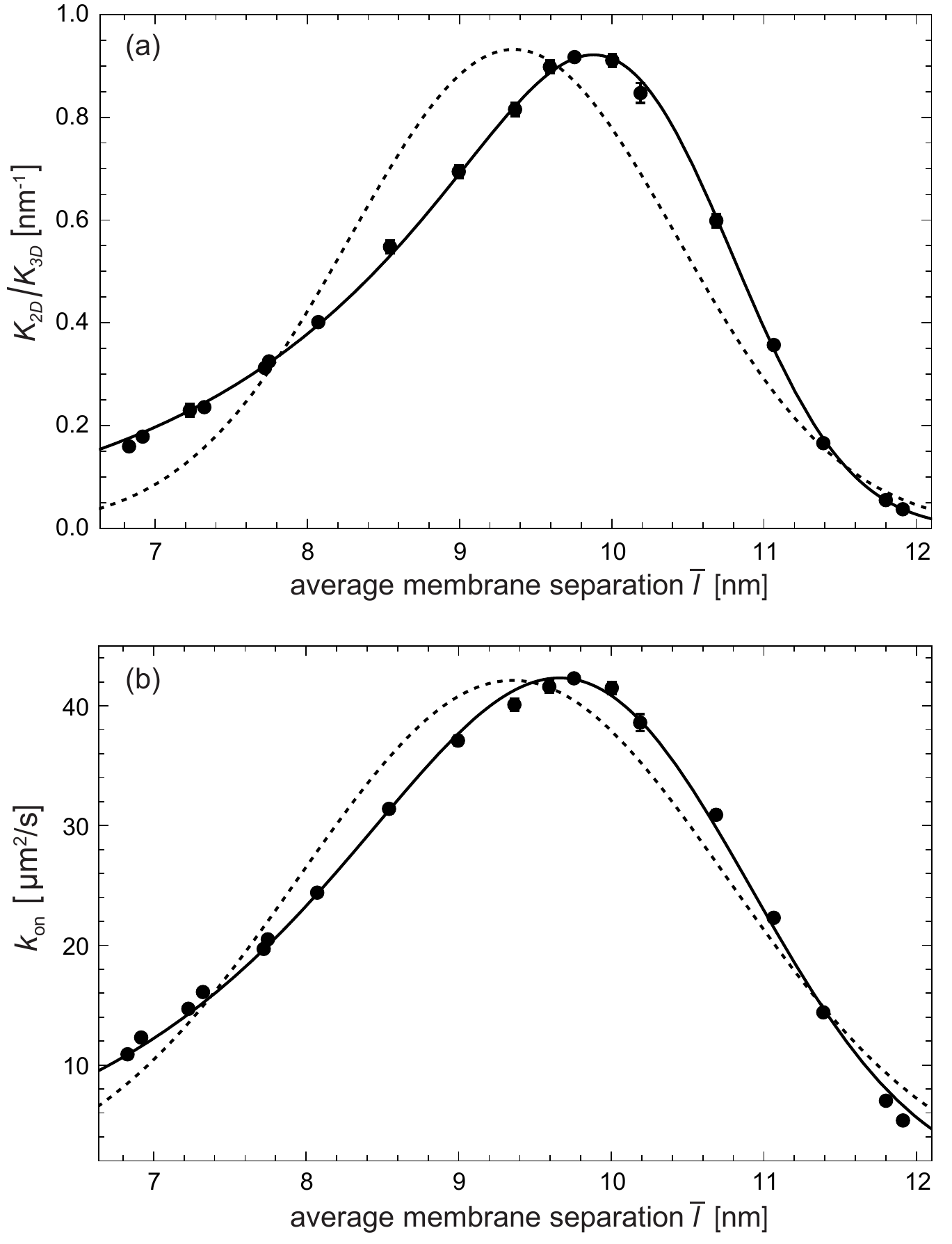}}
\caption{(a) Ratio $K_{\rm 2D}/K_{\rm 3D}$ of the binding constants of lipid-anchored and soluble receptors and ligands and (b) on-rate constant $k_{\rm on}$ of lipid-anchored receptors and ligands versus average membrane separation $\bar{l}$ of two membranes with area $14\times 14$ nm$^2$ and a single lipid-anchored receptor and ligand in our coarse-grained molecular model. The relative membrane roughness is determined by the membrane area in this system and attains the value $\xi_\perp = 0.54 \pm 0.01$ nm. The data points result from MD simulations. The full lines in (a) result from a fit of our general theoretical results for $K_{\rm 2D}/K_{\rm 3D}$ from Eqs.\ (\ref{K2Dav}), (\ref{HRL}), and (\ref{K2Dl-B}) with fit parameters $c_{\rm 2D}=480\pm 20$ nm$^2$,   $L_0 = 10.64\pm 0.02$ nm, and $k_{\rm RL} = 7.2 \pm 0.7$ $k_BT/{\rm nm}^2$ for the anchoring strength $k_a \simeq 2.5$ $k_BT$ of our lipid-anchored receptors and ligands. The full lines in (b) result from a fit of our general theoretical results for $k_{\rm on}$ from Eqs.\ (\ref{HTS}), (\ref{konl}), and (\ref{konav}) with fit parameters $c_{\rm on}=77\pm 4$ $\mu{\rm m}^2/{\rm s}$,   $L_{\rm TS} = 10.63\pm 0.02$ nm, and $k_{\rm TS} = 1.5 \pm 0.2$ $k_BT/{\rm nm}^2$. The dashed lines represent fits to Eqs.\ (\ref{K2D_Gauss}) and (\ref{kon_Gauss}) obtained for the classical Gaussian theory with fit parameters (a) $K_{\rm 2D}^{\rm max} = 146\pm 9$ nm$^2$, $l_K=9.36\pm 0.06$ nm, and $\xi_K = 1.08\pm 0.04$ nm and (b) $k_{\rm on}^{\rm max} = 42.1\pm 1.4$ $\mu{\rm m}^2/{\rm s}$, $l_k=9.35\pm 0.05$ nm, and $\xi_k = 1.41\pm 0.05$ nm.
}
\label{figure-MD-14x14}
\end{figure}

The effective configurational energies (\ref{HRL}) and (\ref{HTS}) describe the bound complex and the transition-state complex of membrane-anchored receptors and ligands as effective harmonic springs that can tilt. In contrast, classical theories describe these complexes as simple harmonic springs \cite{Dembo88,Gao11,Bihr12}.  As functions of the local membrane separation $l$, the binding equilibrium constant $K_{\rm 2D}(l)$ and on-rate constant $k_{\rm on}(l)$ then have a symmetric, Gaussian shape in this classical theory (see Appendix). However, the MC data of Fig.\ \ref{figure-MC-planar} illustrate that the function $K_{\rm 2D}(l)$ is clearly asymmetric, in agreement with Eq.\ (\ref{K2Dl-B}) of our theory. In Fig.\ \ref{figure-MD-14x14}, both our theory (full lines) and the classical theory (dashed lines) are compared to data from MD simulations \cite{Hu15}. In these simulations of our smallest model system with membrane area $14\times 14$ nm$^2$ and a single lipid-anchored receptor and ligand, the average separation $\bar{l}$ of the membranes is varied by varying the number of water beads between the membranes. The relative membrane roughness in this system is determined by the membrane area and attains the value $\xi_\perp\simeq 0.54$ nm. Our theoretical results (full lines) are in good agreement with the MD data. The results for the classical theory (dashed lines) deviate from the data because they do not reflect the asymmetry of $K_{\rm 2D}$ and $k_{\rm on}$ as functions of the average membrane separation $\bar{l}$, which results from the asymmetry of $K_{\rm 2D}(l)$ and $k_{\rm on}(l)$. 

For a relative membrane roughness $\xi_\perp$ that is much larger than the widths $\xi_{\rm RL}$ and $\xi_{\rm TS}$, of the functions $K_{\rm 2D}(l)$ and $k_{\rm on}(l)$,  the distribution $P(l)$ of local membrane separations $l$ is nearly constant over the range of local separations $l$ for which $K_{\rm 2D}(l)$ and $k_{\rm on}(l)$ are not negligibly small. The Eqs.\ (\ref{K2Dav}) and (\ref{konav}) of our theory then simplify to \cite{Xu15,Hu15}
\begin{equation}
K_{\rm 2D} \simeq \frac{\tilde{c}_{\rm 2D}K_{\rm 3D}}{\xi_\perp} \exp\left[-\frac{(\bar{l} -\bar{l}_0)^2}{2\xi_\perp^2}\right]
\label{K2Dlim}
\end{equation}
with $\tilde{c}_{\rm 2D} = k_a A_b/(\sqrt{2\pi k_B T k_{\rm RL}} V_b)$ and
\begin{equation}
k_{\rm on} \simeq 
\frac{\tilde{c}_{\rm on}}{\xi_\perp}\exp\left[-\frac{(\bar{l} -\bar{l}_{\rm TS})^2}{2\xi_\perp^2}\right]
\label{konlim}
\end{equation}
with $\tilde{c}_{\rm on} = c_{\rm on}\pi (k_B T)^{3/2}/(k_a\sqrt{k_{\rm TS}})$ for a Gaussian distribution $P(l)$ of the local membrane separation $l$ (see Eq.\ (\ref{Pl})). Here, $\bar{l}_0$ and $\bar{l}_{\rm TS}$ are the preferred average separations for large roughnesses. For such large roughnesses, the dependence of $K_{\rm 2D}$ and $k_{\rm on}$ on the average separation $\bar{l}$ is dominated by the shape of the distribution $P(l)$, and the asymmetry of $K_{\rm 2D}(l)$ and $k_{\rm on}(l)$ are `averaged out' in Eqs.\ (\ref{K2Dav}) and (\ref{konav}). At the preferred average separations for binding, i.e.\ at the average separations for which the Gaussian functions in Eqs.\ (\ref{K2Dlim}) and (\ref{konlim}) are maximal, the binding constant $K_{\rm 2D}$ and on-rate constant $k_{\rm on}$ are inversely proportional to the relative membrane roughness $\xi_\perp$. 

In our theory, the widths $\xi_{\rm RL}$ and $\xi_{\rm TS}$ of the functions $K_{\rm 2D}(l)$ and $k_{\rm on}(l)$ depends on the anchoring strength $k_a$ of the receptors and ligands, and the preferred lengths and effective spring constants of the bound complex and the transition-state complex \cite{Xu15,Hu15}: 
\begin{align}
\xi_{\rm RL}&\simeq \sqrt{(k_B T/k_{\rm RL}) + (k_BT L_0/2k_a)^2} \\
\xi_{\rm TS}&\simeq \sqrt{(k_B T/k_{\rm TS}) + (k_BT L_{\rm TS}/2k_a)^2}
\end{align}
For the lipid-anchored receptors and ligands of our coarse-grained molecular model, these widths are $\xi_{\rm RL} \simeq 2.1$ nm and $\xi_{\rm TS} \simeq 2.2$ nm. For the transmembrane receptors and ligands, we have $\xi_{\rm RL} \simeq 0.38$ nm and $\xi_{\rm TS} \simeq 0.8$ nm. For the receptors and ligands of our elastic-membrane model, the width $\xi_{\rm RL}$ of the function $K_{\rm 2D}(l)$ ranges between $1.3$ nm and $5.0$ nm, depending on the anchoring strength $k_a$ and complex length $L_0$ of the receptors and ligands. For receptor-ligand complexes of length $L_0 = 40.3$ nm, we have $\xi_{\rm RL} \simeq 5.0$ nm, $2.5$ nm, and $1.3$ nm for the anchoring strengths $k_a = 4 \, k_B T$, $8 \, k_B T$,  and $16 \, k_B T$. For receptors and ligands with anchoring strength $k_a = 8 \, k_B T$, we have $\xi_{\rm RL} \simeq 1.3$ nm, $2.5$ nm, and $3.8$ nm for the complex lengths $L_0 = 20.4$ nm, $40.3$ nm, and $60.3$ nm.

\section{Conclusions and outlook}

The computational model systems and theories reviewed in this article indicate that the relative roughness $\xi_\perp$ of two adhering membranes plays an important role for the binding of membrane-anchored receptors and ligands. For concentrations $[{\rm RL}]$ of receptor-ligand bonds around $100/\mu{\rm m}^{2}$, the relative membrane roughness $\xi_\perp$ obtained from Eq.\ (\ref{relative_roughness_scaling}) is of the same magnitude or larger than the characteristic lengths $\xi_{\rm RL}$ and $\xi_{\rm TS}$ of the receptors and the ligands in our computational model systems, which reflect how strongly the local separation of the membranes is constrained by the receptor-ligand and transition-state complexes. The binding constant $K_{\rm 2D}$ and on-rate constant $k_{\rm on}$ of the receptors and ligands then decreases with increasing relative membrane roughness $\xi_\perp$ in equilibrated membrane adhesion zones in which the average separation $\bar{l}$ of the membranes is close to preferred average separation $\bar{l}_0$ of the receptors and ligands for binding. 

In the next years, experimental model systems of biomembrane adhesion may confirm the effect of the relative membrane roughness  $\xi_\perp$ on the binding constant $K_{\rm 2D}$ of membrane-anchored receptors and ligands. In such model systems, the adhesion of reconstituted membranes is mediated by anchored adhesion proteins  \cite{Albersdoerfer97,Kloboucek99,Maier01,Lorz07,Purrucker07,Reister08,Streicher09,Monzel09,Fenz11,Bihr14,Sackmann14,Schmidt15}, by anchored saccharides \cite{Gourier04,Schneck11}, or  by anchored DNA \cite{Chan07,Beales09,Chung13,Parolini15}. The roughness-dependence of $K_{\rm 2D}$ can be confirmed by demonstrating that $K_{\rm 2D}$ increases with the concentration $[{\rm RL}]$ of bound receptor-ligand complexes, because the relative membrane roughness $\xi_\perp$ decreases with increasing bond concentration $[{\rm RL}]$. Measuring the relative membrane roughness requires a spatial resolution in the nanometer range both in the directions parallel and perpendicular to the membranes, which is beyond the scope of current optical methods used to probe membrane shape fluctuations \cite{Pierres08,Monzel15}. However, the relative membrane roughness can be measured in neutron scattering experiments on stacks of oriented membranes that interact via anchored molecules \cite{Schneck11}.

Our general theories for the binding constant $K_{\rm 2D}$ and binding kinetics of membrane-anchored molecules reviewed in this article are in good agreement with simulation data for our computational model systems. These theories identify characteristic properties of the receptor and ligand molecules and of the apposing membranes that determine the binding equilibrium and kinetics. In the general Eqs.\ (\ref{K2Dav}) and (\ref{konav}), the molecular properties of the receptors and ligands, including their membrane anchoring, are reflected in the functions $K_{\rm 2D}(l)$ and $k_{\rm on}(l)$, and the properties of the membranes are reflected in the distribution $P(l)$ of the local membrane separation $l$. The distribution $P(l)$ has the Gaussian shape (\ref{Pl}) with the average membrane separation $\bar{l}$ and relative membrane roughness $\xi_\perp$ as characteristic lengths if the adhesion is dominated by a single type of receptors and ligands \cite{Hu15,Xu15}. In our detailed theories for $K_{\rm 2D}(l)$ and $k_{\rm on}(l)$ reviewed in Section IV, the receptor-ligand complex and the transition-complex are described as elastic springs that can tilt, which results in asymmetric, non-Gaussian functions $K_{\rm 2D}(l)$ for $k_{\rm on}(l)$. Our theoretical results for the ratio of the binding constants $K_{\rm 2D}$ and $K_{\rm 3D}$ of membrane-anchored and soluble receptors and ligands agree with MC data without any fit parameters (see Figs.\ 2 and 3), which indicates that our theory captures the essential features of the `dimensionality reduction' from 3D to 2D due to membrane anchoring, for both planar and fluctuating membranes. Other theories concern the binding of receptors and ligands anchored to essentially planar membranes \cite{Wu11,Wu13}, the binding of DNA immobilized on apposing nanoparticle surfaces \cite{Leunissen11,Varilly12}, or the binding of flexible receptor and ligand polymers \cite{Jeppesen01,Moreira04,Moore06,Zhang07}.

\appendix

\section*{Appendix: Gaussian theory for mem\-brane-anchored receptors and ligands}

In classical theories \cite{Dembo88,Gao11,Bihr12}, the effective configurational energies $H_{\rm RL}$ and $H_{\rm TS}$ of membrane-anchored receptor-ligand and transition-state complexes depend only on the membrane separation $l$. In harmonic approximation, such effective configurational energies lead to Gaussian functions
\begin{align}
K_{\rm 2D}(l) &= K_{\rm 2D}^{\max}\exp\left[-(l-l_K)^2/2\xi_K^2\right]
\label{K2Dl_Gauss}
 \\
k_{\rm on}(l)  &= k_{\rm on}^{\max}\exp\left[-(l-l_k)^2/2\xi_k^2\right]  
\label{konl_Gauss}
\end{align}
Here, $\xi_K$ and $\xi_k$ are the widths of the functions $K_{\rm 2D}(l)$ and $k_{\rm on}(l)$. For a Gaussian distribution $P(l)$ of the local membrane separation as in Eq.\ (\ref{Pl}), the averages over all local separation $l$ in Eqs.\ (\ref{K2Dav}) and (\ref{konav}) can be calculated explicitly, which leads to 
\begin{align}
K_{\rm 2D} &= \frac{K_{\rm 2D}^{\max}\xi_K}{\sqrt{\xi_\perp^2 + \xi_K^2}}\exp\left[-\frac{(\bar l - l_K)^2}{2(\xi_\perp^2+\xi_K^2)}\right] 
\label{K2D_Gauss}\\
k_{\rm on}  &= \frac{k_{\rm on}^{\max}\xi_k}{\sqrt{\xi_\perp^2+\xi_k^2}}\exp\left[-\frac{(\bar l - l_k)^2}{2(\xi_\perp^2+\xi_k^2)}\right]
\label{kon_Gauss}
\end{align}
From these two equations, we obtain the off-rate constant
\begin{equation}
k_{\rm off} \simeq k_{\rm off}^{\rm min} \exp\left[\frac{(\bar l - l_K)^2}{2(\xi_\perp^2+\xi_K^2)} - \frac{(\bar l - l_k)^2}{2(\xi_\perp^2+\xi_k^2)}\right]
\label{koff_gen}
\end{equation}
with $k_{\rm off}^{\rm min} = (k_{\rm on}^{\max}\xi_k/K^{\max} \xi_K)\sqrt{(\xi_\perp^2 + \xi_K^2)/(\xi_\perp^2+\xi_k^2)}$. Related expressions for averages at fixed membrane locations in the special case $\xi_k = \xi_K$ have been derived by Bihr et al.\ \cite{Bihr12}.

The dependence of the off-rate constant $k_{\rm off}$ on the average membrane separation $\bar l$ can be understood from the first and second derivative of $k_{\rm off}$ with respect to $\bar l$. The first derivative $d\, k_{\rm off}/ d\, \bar l$ vanishes at the average membrane separation
\begin{equation}
\bar{l}_{k^\prime} = \frac{l_K (\xi_k^2+\xi_\perp^2) - l_k(\xi_K^2 + \xi_\perp^2)}{\xi_k^2 - \xi_K^2}
\end{equation}
The value of the second derivative at $d^2 k_{\rm off}/ d\, \bar{l}^2$ at this membrane separation is positive for $\xi_k > \xi_K$, and negative for $\xi_k < \xi_K$. As a function of $\bar l$, the off-rate constant thus exhibits a minimum at $\bar{l} = \bar{l}_{k^\prime}$ for $\xi_k > \xi_K$, and a maximum for $\xi_k < \xi_K$. Depending on the values of $l_K$, $l_k$, $\xi_K$,  $\xi_k$, and $\xi_\perp$, the location $\bar{l}_{k^\prime}$ of this minimum or maximum can adopt values that differ strongly from the locations $l_K$ and $l_k$ of the maxima of the Gaussian functions (\ref{K2Dl_Gauss}) and (\ref{konl_Gauss}). Negative values of $\bar{l}_{k^\prime}$ imply that the off-rate constant $k_{\rm off}$ is monotonously increasing at positive average separations $\bar{l}$ for $\xi_k > \xi_K$, and monotonously decreasing at such average separations for $\xi_k < \xi_K$. Because the membranes cannot intersect, the average separation $\bar{l}$ of the membranes does not attain negative values.


\begin{thebibliography}{100}

\bibitem{Dustin01}
M.~L. Dustin, S.~K. Bromley, M.~M. Davis, and C.~Zhu, ``Identification of self
  through two-dimensional chemistry and synapses,'' {\em Annu. Rev. Cell Dev.
  Biol.}, {\bf 17}, 133--157, 2001.

\bibitem{Orsello01}
C.~E. Orsello, D.~A. Lauffenburger, and D.~A. Hammer, ``Molecular properties in
  cell adhesion: a physical and engineering perspective,'' {\em Trends
  Biotechnol.}, {\bf 19}, 310--316, 2001.

\bibitem{Krobath09}
H.~Krobath, B.~Rozycki, R.~Lipowsky, and T.~R. Weikl, ``Binding cooperativity
  of membrane adhesion receptors,'' {\em Soft Matter}, {\bf 5}, 3354--3361,
  2009.

\bibitem{Wu11}
Y.~Wu, J.~Vendome, L.~Shapiro, A.~Ben-Shaul, and B.~Honig, ``Transforming
  binding affinities from three dimensions to two with application to cadherin
  clustering,'' {\em Nature}, {\bf 475}, 510--513, 2011.

\bibitem{Leckband12}
D.~Leckband and S.~Sivasankar, ``Cadherin recognition and adhesion,'' {\em
  Curr. Opin. Cell. Biol.}, {\bf 24}, 620--627, 2012.

\bibitem{Zarnitsyna12}
V.~Zarnitsyna and C.~Zhu, ``T cell triggering: insights from {2D} kinetics
  analysis of molecular interactions,'' {\em Phys. Biol.}, {\bf 9}, 045005, 2012.

\bibitem{Hu13}
J.~Hu, R.~Lipowsky, and T.~R. Weikl, ``Binding constants of membrane-anchored
  receptors and ligands depend strongly on the nanoscale roughness of
  membranes,'' {\em Proc. Natl. Acad. Sci. USA}, {\bf 110}, 15283--15288, 2013.

\bibitem{Wu13}
Y.~Wu, B.~Honig, and A.~Ben-Shaul, ``Theory and simulations of adhesion
  receptor dimerization on membrane surfaces,'' {\em Biophys. J.}, {\bf 104}, 1221--1229, 2013.

\bibitem{Xu15}
G.-K. Xu, J.~Hu, R.~Lipowsky, and T.~R. Weikl, ``Binding constants of
  membrane-anchored receptors and ligands: A general theory corroborated {Monte
  Carlo} simulations,'' {\em J. Chem. Phys.}, {\bf 143}, 243136, 2015.

\bibitem{Hu15}
J.~Hu, G.-K. Xu, R.~Lipowsky, and T.~R. Weikl, ``Binding kinetics of
  membrane-anchored receptors and ligands: Molecular dynamics simulations and
  theory,'' {\em J. Chem. Phys}, {\bf 143}, 243137, 2015.

\bibitem{Schuck97}
P.~Schuck, ``Use of surface plasmon resonance to probe the equilibrium and
  dynamic aspects of interactions between biological macromolecules,'' {\em
  Annu. Rev. Biophys. Biomol. Struct.}, {\bf 26}, 541--566, 1997.

\bibitem{Rich00}
R.~L. Rich and D.~G. Myszka, ``Advances in surface plasmon resonance biosensor
  analysis,'' {\em Curr. Opin. Biotechnol.}, {\bf 11}, 54--61, 2000.

\bibitem{McDonnell01}
J.~M. McDonnell, ``Surface plasmon resonance: towards an understanding of the
  mechanisms of biological molecular recognition,'' {\em Curr. Opin. Chem.
  Biol.}, {\bf 5}, 572--577, 2001.

\bibitem{Bell84}
G.~I. Bell, M.~Dembo, and P.~Bongrand, ``{Cell adhesion. Competition between
  nonspecific repulsion and specific bonding},'' {\em Biophys. J.}, {\bf 45}, 1051--1064, 1984.

\bibitem{Dustin96}
M.~L. Dustin, L.~M. Ferguson, P.~Y. Chan, T.~A. Springer, and D.~E. Golan,
  ``{Visualization of CD2 interaction with LFA-3 and determination of the
  two-dimensional dissociation constant for adhesion receptors in a contact
  area},'' {\em J. Cell. Biol.}, {\bf 132}, 465--474, 1996.

\bibitem{Dustin97}
M.~L. Dustin, D.~E. Golan, D.~M. Zhu, J.~M. Miller, W.~Meier, E.~A. Davies, and
  P.~A. van~der Merwe, ``{Low affinity interaction of human or rat T cell
  adhesion molecule CD2 with its ligand aligns adhering membranes to achieve
  high physiological affinity},'' {\em J. Biol. Chem.}, {\bf 272}, 30889--30898, 1997.

\bibitem{Zhu07}
D.-M. Zhu, M.~L. Dustin, C.~W. Cairo, and D.~E. Golan, ``Analysis of
  two-dimensional dissociation constant of laterally mobile cell adhesion
  molecules,'' {\em Biophys. J.}, {\bf 92}, 1022--1034, 2007.

\bibitem{Tolentino08}
T.~P. Tolentino, J.~Wu, V.~I. Zarnitsyna, Y.~Fang, M.~L. Dustin, and C.~Zhu,
  ``{Measuring diffusion and binding kinetics by contact area FRAP},'' {\em
  Biophys. J.}, {\bf 95}, 920--930, 2008.

\bibitem{Huppa10}
J.~B. Huppa, M.~Axmann, M.~A. M{\"o}rtelmaier, B.~F. Lillemeier, E.~W. Newell,
  M.~Brameshuber, L.~O. Klein, G.~J. Sch\"utz, and M.~M. Davis,
  ``{TCR-peptide-MHC interactions in situ show accelerated kinetics and
  increased affinity},'' {\em Nature}, {\bf 463}, 963--967, 2010.

\bibitem{Axmann12}
M.~Axmann, J.~B. Huppa, M.~M. Davis, and G.~J. Sch\"utz, ``Determination of
  interaction kinetics between the {T} cell receptor and peptide-loaded {MHC}
  class {II} via single-molecule diffusion measurements,'' {\em Biophys. J.},
  {\bf 103}, L17--L19, 2012.

\bibitem{ODonoghue13}
G.~P. O'Donoghue, R.~M. Pielak, A.~A. Smoligovets, J.~J. Lin, and J.~T. Groves,
  ``{Direct single molecule measurement of TCR triggering by agonist pMHC in
  living primary T cells},'' {\em Elife}, {\bf 2}, e00778, 2013.

\bibitem{Kaplanski93}
G.~Kaplanski, C.~Farnarier, O.~Tissot, A.~Pierres, A.~M. Benoliel, M.~C.
  Alessi, S.~Kaplanski, and P.~Bongrand, ``{Granulocyte-endothelium initial
  adhesion. Analysis of transient binding events mediated by E-selectin in a
  laminar shear flow.},'' {\em Biophys. J.}, {\bf 64}, 1922--1933, 1993.

\bibitem{Alon95}
R.~Alon, D.~A. Hammer, and T.~A. Springer, ``{Lifetime of the
  P-selectin-carbohydrate bond and its response to tensile force in
  hydrodynamic flow},'' {\em Nature}, {\bf 374}, 539--542, 1995.

\bibitem{Piper98}
J.~W. Piper, R.~A. Swerlick, and C.~Zhu, ``Determining force dependence of
  two-dimensional receptor-ligand binding affinity by centrifugation,'' {\em
  Biophys. J.}, {\bf 74}, 492--513, 1998.

\bibitem{Chesla98}
S.~E. Chesla, P.~Selvaraj, and C.~Zhu, ``{Measuring two-dimensional
  receptor-ligand binding kinetics by micropipette},'' {\em Biophys. J.},
  {\bf 75}, 1553--1572, 1998.

\bibitem{Merkel99}
R.~Merkel, P.~Nassoy, A.~Leung, K.~Ritchie, and E.~Evans, ``Energy landscapes
  of receptor-ligand bonds explored with dynamic force spectroscopy,'' {\em
  Nature}, {\bf 397}, 50--53, 1999.

\bibitem{Williams01}
T.~E. Williams, S.~Nagarajan, P.~Selvaraj, and C.~Zhu, ``Quantifying the impact
  of membrane microtopology on effective two-dimensional affinity.,'' {\em J.
  Biol. Chem.}, {\bf 276}, 13283--13288, 2001.

\bibitem{Chen08}
W.~Chen, E.~A. Evans, R.~P. McEver, and C.~Zhu, ``Monitoring receptor-ligand
  interactions between surfaces by thermal fluctuations,'' {\em Biophys. J.},
  {\bf 94}, 694--701, 2008.

\bibitem{Chien08}
Y.-H. Chien, N.~Jiang, F.~Li, F.~Zhang, C.~Zhu, and D.~Leckband, ``Two stage
  cadherin kinetics require multiple extracellular domains but not the
  cytoplasmic region,'' {\em J. Biol. Chem.}, {\bf 283}, 1848--1856, 2008.

\bibitem{Huang10}
J.~Huang, V.~I. Zarnitsyna, B.~Liu, L.~J. Edwards, N.~Jiang, B.~D. Evavold, and
  C.~Zhu, ``{The kinetics of two-dimensional TCR and pMHC interactions
  determine T-cell responsiveness},'' {\em Nature}, {\bf 464}, 932--936, 2010.

\bibitem{Liu14}
B.~Liu, W.~Chen, B.~D. Evavold, and C.~Zhu, ``{Accumulation of dynamic catch
  bonds between TCR and agonist peptide-MHC triggers T cell signaling},'' {\em
  Cell}, {\bf 157}, 357--368, 2014.

\bibitem{Bihr12}
T.~Bihr, U.~Seifert, and A.-S. Smith, ``Nucleation of ligand-receptor domains
  in membrane adhesion,'' {\em Phys. Rev. Lett.}, {\bf 109}, 258101, 2012.

\bibitem{Krobath07}
H.~Krobath, G.~J. Sch\"utz, R.~Lipowsky, and T.~R. Weikl, ``{Lateral diffusion
  of receptor-ligand bonds in membrane adhesion zones: Effect of thermal
  membrane roughness},'' {\em Europhys. Lett.}, {\bf 78}, 38003, 2007.

\bibitem{Nagle13}
J.~F. Nagle, ``Introductory lecture: Basic quantities in model biomembranes,''
  {\em Faraday Discuss.}, {\bf 161}, 11--29, 2013.

\bibitem{Dimova14}
R.~Dimova, ``Recent developments in the field of bending rigidity measurements
  on membranes,'' {\em Adv. Colloid Interface Sci.}, {\bf 208}, 225--234, 2014.

\bibitem{Pontes13}
B.~Pontes, Y.~Ayala, A.~C.~C. Fonseca, L.~F. Romao, R.~F. Amaral, L.~T.
  Salgado, F.~R. Lima, M.~Farina, N.~B. Viana, V.~Moura-Neto, and H.~M.
  Nussenzveig, ``Membrane elastic properties and cell function,'' {\em {PLoS
  One}}, {\bf 8}, 2013.

\bibitem{Betz09}
T.~Betz, M.~Lenz, J.-F. Joanny, and C.~Sykes, ``Atp-dependent mechanics of red
  blood cells,'' {\em Proc. Natl. Acad. Sci. USA}, {\bf 106}, 15320--15325, 2009.

\bibitem{Zamir01}
E.~Zamir and B.~Geiger, ``Molecular complexity and dynamics of cell-matrix
  adhesions,'' {\em J. Cell. Sci.}, {\bf 114}, 3583--90, 2001.

\bibitem{Leckband14}
D.~E. Leckband and J.~de~Rooij, ``Cadherin adhesion and mechanotransduction,''
  {\em Annu. Rev. Cell Dev. Biol.}, {\bf 30}, 291--315, 2014.

\bibitem{Biswas15}
K.~H. Biswas, K.~L. Hartman, C.-h. Yu, O.~J. Harrison, H.~Song, A.~W. Smith,
  W.~Y.~C. Huang, W.-C. Lin, Z.~Guo, A.~Padmanabhan, S.~M. Troyanovsky, M.~L.
  Dustin, L.~Shapiro, B.~Honig, R.~Zaidel-Bar, and J.~T. Groves, ``E-cadherin
  junction formation involves an active kinetic nucleation process,'' {\em
  Proc. Natl. Acad. Sci. USA}, {\bf 112}, 10932--10937, 2015.

\bibitem{Yap15}
A.~S. Yap, G.~A. Gomez, and R.~G. Parton, ``Adherens junctions revisualized:
  Organizing cadherins as nanoassemblies,'' {\em Dev. Cell}, {\bf 35}, 12--20, 2015.

\bibitem{Goetz98}
R.~Goetz and R.~Lipowsky, ``Computer simulations of bilayer membranes:
  Self-assembly and interfacial tension,'' {\em J. Chem. Phys.}, {\bf 108}, 7397--7409, 1998.

\bibitem{Shelley01}
J.~C. Shelley, M.~Y. Shelley, R.~C. Reeder, S.~Bandyopadhyay, and M.~L. Klein,
  ``A coarse grain model for phospholipid simulations,'' {\em J. Phys. Chem.
  B}, {\bf 105}, 4464--4470, 2001.

\bibitem{Marrink04}
S.~J. Marrink, A.~H. de~Vries, and A.~E. Mark, ``Coarse grained model for
  semiquantitative lipid simulations,'' {\em J. Phys. Chem. B}, {\bf 108}, 750--760, 2004.

\bibitem{Shih06}
A.~Y. Shih, A.~Arkhipov, P.~L. Freddolino, and K.~Schulten, ``Coarse grained
  protein-lipid model with application to lipoprotein particles,'' {\em J.
  Phys. Chem. B.}, {\bf 110}, 3674--3684, 2006.

\bibitem{Marrink03}
S.~J. Marrink and A.~E. Mark, ``The mechanism of vesicle fusion as revealed by
  molecular dynamics simulations,'' {\em J. Am. Chem. Soc.}, {\bf 125}, 11144--11145, 2003.

\bibitem{Shillcock05}
J.~C. Shillcock and R.~Lipowsky, ``Tension-induced fusion of bilayer membranes
  and vesicles,'' {\em Nat. Mater.}, {\bf 4}, 225--8, 2005.

\bibitem{Grafmuller07}
A.~Grafm{\"u}ller, J.~Shillcock, and R.~Lipowsky, ``Pathway of membrane fusion
  with two tension-dependent energy barriers,'' {\em Phys. Rev. Lett.},
  {\bf 98}, 218101, 2007.

\bibitem{Grafmuller09}
A.~Grafm{\"u}ller, J.~Shillcock, and R.~Lipowsky, ``The fusion of membranes and
  vesicles: pathway and energy barriers from dissipative particle dynamics,''
  {\em Biophys. J.}, {\bf 96}, 2658--2675, 2009.

\bibitem{Smirnova10}
Y.~G. Smirnova, S.-J. Marrink, R.~Lipowsky, and V.~Knecht, ``Solvent-exposed
  tails as prestalk transition states for membrane fusion at low hydration,''
  {\em J. Am. Chem. Soc.}, {\bf 132}, 6710--6718, 2010.

\bibitem{Risselada11}
H.~J. Risselada, C.~Kutzner, and H.~Grubmueller, ``Caught in the act:
  Visualization of {SNARE}-mediated fusion events in molecular detail,'' {\em
  ChemBioChem}, {\bf 12}, 1049--1055, 2011.

\bibitem{Illya06}
G.~Illya, R.~Lipowsky, and J.~C. Shillcock, ``Two-component membrane material
  properties and domain formation from dissipative particle dynamics,'' {\em J.
  Chem. Phys.}, {\bf 125},  2006.

\bibitem{Risselada08}
H.~J. Risselada and S.~J. Marrink, ``The molecular face of lipid rafts in model
  membranes,'' {\em Proc. Natl. Acad. Sci. USA}, {\bf 105}, 17367--17372, 2008.

\bibitem{Meyer10}
F.~J.-M. de~Meyer, A.~Benjamini, J.~M. Rodgers, Y.~Misteli, and B.~Smit,
  ``Molecular simulation of the {DMPC}-cholesterol phase diagram,'' {\em J.
  Phys. Chem. B}, {\bf 114}, 10451--10461, 2010.

\bibitem{Apajalahti10}
T.~Apajalahti, P.~Niemela, P.~N. Govindan, M.~S. Miettinen, E.~Salonen, S.-J.
  Marrink, and I.~Vattulainen, ``Concerted diffusion of lipids in raft-like
  membranes,'' {\em Faraday Discuss.}, {\bf 144}, 411--430, 2010.

\bibitem{Bennett13}
W.~F.~D. Bennett and D.~P. Tieleman, ``Computer simulations of lipid membrane
  domains,'' {\em Biochim. Biophys. Acta-Biomembr.}, {\bf 1828}, 1765--1776, 2013.

\bibitem{Gambin06}
Y.~Gambin, R.~Lopez-Esparza, M.~Reffay, E.~Sierecki, N.~S. Gov, M.~Genest,
  R.~S. Hodges, and W.~Urbach, ``Lateral mobility of proteins in liquid
  membranes revisited,'' {\em Proc. Natl. Acad. Sci. USA}, {\bf 103}, 2098--2102, 2006.

\bibitem{Guigas06}
G.~Guigas and M.~Weiss, ``Size-dependent diffusion of membrane inclusions,''
  {\em Biophys. J.}, {\bf 91}, 2393--2398, 2006.

\bibitem{Reynwar07}
B.~J. Reynwar, G.~Illya, V.~A. Harmandaris, M.~M. M\"uller, K.~Kremer, and
  M.~Deserno, ``Aggregation and vesiculation of membrane proteins by
  curvature-mediated interactions,'' {\em Nature}, {\bf 447}, 461--464,
  2007.

\bibitem{Arkhipov08}
A.~Arkhipov, Y.~Yin, and K.~Schulten, ``Four-scale description of membrane
  sculpting by {BAR} domains,'' {\em Biophys. J.}, {\bf 95}, 2806--2821, 2008.

\bibitem{Simunovic13}
M.~Simunovic, A.~Srivastava, and G.~A. Voth, ``Linear aggregation of proteins
  on the membrane as a prelude to membrane remodeling,'' {\em Proc. Natl. Acad.
  Sci. U. S. A.}, {\bf 110}, 20396--20401, 2013.

\bibitem{Lipowsky96}
R.~Lipowsky, ``Adhesion of membranes via anchored stickers,'' {\em Phys. Rev.
  Lett.}, {\bf 77}, 1652--1655, 1996.

\bibitem{Weikl01}
T.~R. Weikl and R.~Lipowsky, ``Adhesion-induced phase behavior of
  multicomponent membranes,'' {\em Phys. Rev. E.}, {\bf 64}, 011903, 2001.

\bibitem{Weikl02a}
T.~R. Weikl, J.~T. Groves, and R.~Lipowsky, ``Pattern formation during adhesion
  of multicomponent membranes,'' {\em Europhys. Lett.}, {\bf 59}, 916--922,
  2002.

\bibitem{Weikl04}
T.~R. Weikl and R.~Lipowsky, ``{Pattern formation during T-cell adhesion},''
  {\em Biophys. J.}, {\bf 87}, 3665--3678, 2004.

\bibitem{Asfaw06}
M.~Asfaw, B.~Rozycki, R.~Lipowsky, and T.~R. Weikl, ``Membrane adhesion via
  competing receptor/ligand bonds,'' {\em Europhys. Lett.}, {\bf 76}, 703--709, 2006.

\bibitem{Tsourkas07}
P.~K. Tsourkas, N.~Baumgarth, S.~I. Simon, and S.~Raychaudhuri, ``{Mechanisms
  of B-cell synapse formation predicted by Monte Carlo simulation},'' {\em
  Biophys. J.}, {\bf 92}, 4196--4208, 2007.

\bibitem{Reister08}
E.~Reister-Gottfried, K.~Sengupta, B.~Lorz, E.~Sackmann, U.~Seifert, and A.~S.
  Smith, ``Dynamics of specific vesicle-substrate adhesion: From local events
  to global dynamics,'' {\em Phys. Rev. Lett.}, {\bf 101}, 208103, 2008.

\bibitem{Paszek09}
M.~J. Paszek, D.~Boettiger, V.~M. Weaver, and D.~A. Hammer, ``Integrin
  clustering is driven by mechanical resistance from the glycocalyx and the
  substrate,'' {\em PLoS Comput. Biol.}, {\bf 5}, e1000604, 2009.

\bibitem{Komura00}
S.~Komura and D.~Andelman, ``Adhesion-induced lateral phase separation in
  membranes,'' {\em Eur. Phys. J. E}, {\bf 3}, 259--271, 2000.

\bibitem{Bruinsma00}
R.~Bruinsma, A.~Behrisch, and E.~Sackmann, ``Adhesive switching of membranes:
  experiment and theory,'' {\em Phys. Rev. E}, {\bf 61}, 4253--4267, 2000.

\bibitem{Qi01}
S.~Y. Qi, J.~T. Groves, and A.~K. Chakraborty, ``Synaptic pattern formation
  during cellular recognition,'' {\em Proc. Natl. Acad. Sci. USA}, {\bf 98}, 6548--6553, 2001.

\bibitem{Chen03}
H.-Y. Chen, ``Adhesion-induced phase separation of multiple species of membrane
  junctions,'' {\em Phys. Rev. E}, {\bf 67}, 031919, 2003.

\bibitem{Raychaudhuri03}
S.~Raychaudhuri, A.~K. Chakraborty, and M.~Kardar, ``Effective membrane model
  of the immunological synapse,'' {\em Phys. Rev. Lett.}, {\bf 91}, 208101,
  2003.

\bibitem{Coombs04}
D.~Coombs, M.~Dembo, C.~Wofsy, and B.~Goldstein, ``Equilibrium thermodynamics
  of cell-cell adhesion mediated by multiple ligand-receptor pairs,'' {\em
  Biophys. J.}, {\bf 86}, 1408--1423, 2004.

\bibitem{Shenoy05}
V.~B. Shenoy and L.~B. Freund, ``Growth and shape stability of a biological
  membrane adhesion complex in the diffusion-mediated regime,'' {\em Proc.
  Natl. Acad. Sci. USA}, {\bf 102}, 3213--3218, 2005.

\bibitem{Wu06}
J.-Y. Wu and H.-Y. Chen, ``Membrane-adhesion-induced phase separation of two
  species of junctions,'' {\em Phys. Rev. E}, {\bf 73}, 011914, 2006.

\bibitem{Zhang08a}
C.-Z. Zhang and Z.-G. Wang, ``Nucleation of membrane adhesions,'' {\em Phys.
  Rev. E}, {\bf 77}, 021906, 2008.

\bibitem{Atilgan09}
E.~Atilgan and B.~Ovryn, ``Nucleation and growth of integrin adhesions,'' {\em
  Biophys J}, {\bf 96}, 3555--3572, 2009.

\bibitem{Zuckerman95}
D.~Zuckerman and R.~Bruinsma, ``Statistical mechanics of membrane adhesion by
  reversible molecular bonds,'' {\em Phys. Rev. Lett.}, {\bf 74}, 3900--3903, 1995.

\bibitem{Speck10}
T.~Speck, E.~Reister, and U.~Seifert, ``Specific adhesion of membranes: Mapping
  to an effective bond lattice gas,'' {\em Phys. Rev. E}, {\bf 82},  2010.

\bibitem{Weil10}
N.~Weil and O.~Farago, ``Entropy-driven aggregation of adhesion sites of
  supported membranes,'' {\em Eur. Phys. J. E}, {\bf 33}, 81--87, 2010.

\bibitem{Dharan15}
N.~Dharan and O.~Farago, ``Formation of adhesion domains in stressed and
  confined membranes,'' {\em Soft Matter}, {\bf 11}, 3780--3785,
  2015.

\bibitem{Israelachvili92}
J.~N. Israelachvili, {\em Intermolecular and surface forces, 2nd ed.}
\newblock {Academic Press}, 1992.

\bibitem{Bayas07}
M.~V. Bayas, A.~Kearney, A.~Avramovic, P.~A. van~der Merwe, and D.~E. Leckband,
  ``{Impact of salt bridges on the equilibrium binding and adhesion of human
  CD2 and CD58},'' {\em J. Biol. Chem.}, {\bf 282}, 5589--5596, 2007.

\bibitem{Robert11}
P.~Robert, A.~Nicolas, S.~Aranda-Espinoza, P.~Bongrand, and L.~Limozin,
  ``Minimal encounter time and separation determine ligand-receptor binding in
  cell adhesion,'' {\em Biophys. J.}, {\bf 100}, 2642--2651, 2011.

\bibitem{Wang12}
J.-h. Wang and E.~L. Reinherz, ``The structural basis of $\alpha\beta$
  {T}-lineage immune recognition: {TCR} docking topologies,
  mechanotransduction, and co-receptor function,'' {\em Immunol. Rev.},
  {\bf 250}, 102--119, 2012.

\bibitem{Dembo88}
M.~Dembo, D.~C. Torney, K.~Saxman, and D.~Hammer, ``The reaction-limited
  kinetics of membrane-to-surface adhesion and detachment,'' {\em Proc. R. Soc.
  Lond. B}, {\bf  234}, 55--83, 1988.

\bibitem{Gao11}
H.~Gao, J.~Qian, and B.~Chen, ``Probing mechanical principles of focal contacts
  in cell-matrix adhesion with a coupled stochastic-elastic modelling
  framework,'' {\em J. R. Soc. Interface}, {\bf 8}, 1217--1232, 2011.

\bibitem{Albersdoerfer97}
A.~Albersd\"orfer, T.~Feder, and E.~Sackmann, ``Adhesion-induced domain
  formation by interplay of long-range repulsion and short-range attraction
  force: A model membrane study,'' {\em Biophys. J.}, {\bf 73}, 245--257,
1997.

\bibitem{Kloboucek99}
A.~Kloboucek, A.~Behrisch, J.~Faix, and E.~Sackmann, ``{Adhesion-induced
  receptor segregation and adhesion plaque formation: A model membrane
  study},'' {\em Biophys. J.}, {\bf 77}, 2311--2328, 1999.

\bibitem{Maier01}
C.~W. Maier, A.~Behrisch, A.~Kloboucek, D.~A. Simson, and R.~Merkel, ``Specific
  biomembrane adhesion - indirect lateral interactions between bound receptor
  molecules,'' {\em Eur. Phys. J. E}, {\bf 6}, 273--276, 2001.

\bibitem{Lorz07}
B.~G. Lorz, A.-S. Smith, C.~Gege, and E.~Sackmann, ``{Adhesion of giant
  vesicles mediated by weak binding of Sialyl-Lewis(x) to E-selectin in the
  presence of repelling poly(ethylene glycol) molecules},'' {\em Langmuir},
  {\bf 23}, 12293--12300, 2007.

\bibitem{Purrucker07}
O.~Purrucker, S.~Goennenwein, A.~Foertig, R.~Jordan, M.~Rusp, M.~Baermann,
  L.~Moroder, E.~Sackmann, and M.~Tanaka, ``Polymer-tethered membranes as
  quantitative models for the study of integrin-mediated cell adhesion,'' {\em
  Soft Matter}, {\bf 3}, 333--336, 2007.

\bibitem{Streicher09}
P.~Streicher, P.~Nassoy, M.~B{\"a}rmann, A.~Dif, V.~Marchi-Artzner,
  F.~Brochard-Wyart, J.~Spatz, and P.~Bassereau, ``{Integrin reconstituted in
  GUVs: a biomimetic system to study initial steps of cell spreading},'' {\em
  Biochim Biophys Acta}, {\bf 1788}, 2291--2300, 2009.

\bibitem{Monzel09}
C.~Monzel, S.~F. Fenz, R.~Merkel, and K.~Sengupta, ``Probing biomembrane
  dynamics by dual-wavelength reflection interference contrast microscopy,''
  {\em Chemphyschem}, {\bf 10}, 2828--2838, 2009.

\bibitem{Fenz11}
S.~F. Fenz, A.-S. Smith, R.~Merkel, and K.~Sengupta, ``Inter-membrane adhesion
  mediated by mobile linkers: Effect of receptor shortage,'' {\em Soft Matter},
  {\bf 7}, 952--962, 2011.

\bibitem{Bihr14}
T.~Bihr, S.~Fenz, E.~Sackmann, R.~Merkel, U.~Seifert, K.~Sengupta, and A.-S.
  Smith, ``Association rates of membrane-coupled cell adhesion molecules,''
  {\em Biophys. J.}, {\bf 107}, L33--L36, 2014.

\bibitem{Sackmann14}
E.~Sackmann and A.-S. Smith, ``Physics of cell adhesion: some lessons from
  cell-mimetic systems,'' {\em Soft Matter}, {\bf 10}, 1644--1659,
  2014.

\bibitem{Schmidt15}
D.~Schmidt, T.~Bihr, S.~Fenz, R.~Merkel, U.~Seifert, K.~Sengupta, and A.-S.
  Smith, ``Crowding of receptors induces ring-like adhesions in model
  membranes,'' {\em Biochim. Biophys. Acta-Mol. Cell Res.}, {\bf 1853}, 2984--2991, 2015.

\bibitem{Gourier04}
C.~Gourier, F.~Pincet, E.~Perez, Y.~Zhang, J.-M. Mallet, and P.~Sinay,
  ``Specific and non specific interactions involving {LeX} determinant
  quantified by lipid vesicle micromanipulation,'' {\em Glycoconjugate J.},
  {\bf 21}, 165--174, 2004.

\bibitem{Schneck11}
E.~Schneck, B.~Deme, C.~Gege, and M.~Tanaka, ``Membrane adhesion via homophilic
  saccharide-saccharide interactions investigated by neutron scattering,'' {\em
  Biophys. J.}, {\bf 100}, 2151--2159, 2011.

\bibitem{Chan07}
Y.-H.~M. Chan, P.~Lenz, and S.~G. Boxer, ``Kinetics of {DNA}-mediated docking
  reactions between vesicles tethered to supported lipid bilayers,'' {\em Proc.
  Natl. Acad. Sci. USA}, {\bf 104}, 18913--8, 2007.

\bibitem{Beales09}
P.~A. Beales and T.~K. Vanderlick, ``{DNA} as membrane-bound ligand-receptor
  pairs: Duplex stability is tuned by intermembrane forces,'' {\em Biophys.
  J.}, {\bf 96}, 1554--1565,  2009.

\bibitem{Chung13}
M.~Chung, B.~J. Koo, and S.~G. Boxer, ``Formation and analysis of topographical
  domains between lipid membranes tethered by {DNA} hybrids of different
  lengths,'' {\em Faraday Discuss.}, {\bf 161}, 333--345,
  2013.

\bibitem{Parolini15}
L.~Parolini, B.~M. Mognetti, J.~Kotar, E.~Eiser, P.~Cicuta, and L.~Di~Michele,
  ``Volume and porosity thermal regulation in lipid mesophases by coupling
  mobile ligands to soft membranes,'' {\em Nat. Commun.}, {\bf 6}, 2015.

\bibitem{Pierres08}
A.~Pierres, A.-M. Benoliel, D.~Touchard, and P.~Bongrand, ``How cells tiptoe on
  adhesive surfaces before sticking,'' {\em Biophys J.}, {\bf 94}, 4114--4122, 2008.

\bibitem{Monzel15}
C.~Monzel, D.~Schmidt, C.~Kleusch, D.~Kirchenbuechler, U.~Seifert, A.-S. Smith,
  K.~Sengupta, and R.~Merkel, ``Measuring fast stochastic displacements of
  bio-membranes with dynamic optical displacement spectroscopy,'' {\em Nat.
  Commun.}, {\bf 6},  2015.

\bibitem{Leunissen11}
M.~E. Leunissen and D.~Frenkel, ``Numerical study of {DNA}-functionalized
  microparticles and nanoparticles: explicit pair potentials and their
  implications for phase behavior,'' {\em J. Chem. Phys.}, {\bf 134}, 084702, 2011.

\bibitem{Varilly12}
P.~Varilly, S.~Angioletti-Uberti, B.~M. Mognetti, and D.~Frenkel, ``A general
  theory of {DNA}-mediated and other valence-limited colloidal interactions,''
  {\em J. Chem. Phys.}, {\bf 137}, 094108,  2012.

\bibitem{Jeppesen01}
C.~Jeppesen, J.~Y. Wong, T.~L. Kuhl, J.~N. Israelachvili, N.~Mullah,
  S.~Zalipsky, and C.~M. Marques, ``Impact of polymer tether length on multiple
  ligand-receptor bond formation,'' {\em Science}, {\bf 293}, 465--468,
  2001.

\bibitem{Moreira04}
A.~G. Moreira and C.~M. Marques, ``The role of polymer spacers in specific
  adhesion,'' {\em J. Chem. Phys.}, {\bf 120}, 6229--6237, 2004.

\bibitem{Moore06}
N.~W. Moore and T.~L. Kuhl, ``The role of flexible tethers in multiple
  ligand-receptor bond formation between curved surfaces,'' {\em Biophys. J.},
  {\bf 91}, 1675--1687, 2006.

\bibitem{Zhang07}
C.-Z. Zhang and Z.-G. Wang, ``Polymer-tethered ligand-receptor interactions
  between surfaces ii,'' {\em Langmuir}, {\bf 23}, 13024--13039, 2007.

\end{thebibliography}
\end{document}